%% file: main.tex
\documentclass{article}

\usepackage{PRIMEarxiv}

\usepackage[utf8]{inputenc} 
\usepackage[T1]{fontenc}    
\usepackage{hyperref}       
\usepackage{url}            
\usepackage{booktabs}       
\usepackage{amsfonts}       
\usepackage{nicefrac}       
\usepackage{microtype}      
\usepackage{lipsum}
\usepackage{algpseudocode}
\usepackage{textcomp}
\usepackage{algorithm}
\usepackage{fancyhdr}       
\usepackage{graphicx}       
\usepackage{optidef}
\usepackage{bm}
\usepackage{longtable,tabularx}
\usepackage{float}
\usepackage{amsmath}

\title{Deep neural operators can serve as accurate surrogates for shape optimization: A case study for airfoils
}

\author{
  *Khemraj Shukla, *Vivek Oommen, *Ahmad Peyvan \\
  Brown University, Providence, RI, 02912 \\
 \\
   \And
  *Michael Penwarden \\
  University of Utah, Salt Lake City, UT, 84112 \\
  \AND
  Luis Bravo, Anindya Ghoshal \\
  Weapons and Materials Directorate, U.S. Army Research Laboratory \\
  Aberdeen Proving Ground, MD, 21005 \\
  \\
  \And
  Robert M. Kirby \\
  University of Utah, Salt Lake City, UT, 84112 \\
 \\
  \And
  George Em Karniadakis \\
   Brown University, Providence, RI, 02912 \\
   \\
  \And
  *Equal contribution\\
 \\
}

\begin{document}
\maketitle

\input{abstract}


\section*{Nomenclature}

{\renewcommand\arraystretch{0.75}
\noindent\begin{longtable}{l@{\quad=\quad} l} 

$m$ & Maximum camber in percentage of the chord  \\
$p$ & Position of maximum camber in percentage of the chord \\
$t$ & Maximum thickness of the airfoil in percentage of the chord  \\ 
$\rho$  & non-dimensional density  \\
$u$ & non-dimensional velocity of the fluid in \textit{x} direction \\
$v$ &  non-dimensional velocity of the fluid in \textit{y} direction  \\
$p$ & non-dimensional pressure  \\
$T$ & non-dimensional temperature  \\
$\mathcal{G}$ & output of the DeepONet model  \\
$\theta_b$ & set of all trainable weights and biases of the branch network \\
$\theta_t$ & set of all trainable weights and biases of the trunk network  \\
$N_{\phi}$ & number of basis functions learned by the DeepONet  \\
$\tau_{ij}$ & Components of viscous stress tensors \\
$f$ & generalized input function to the DeepONet \\
$\xi_g$ &  Geometric parameter\\
$c$ & Chord length\\
$Ma$ & Mach number\\
$\xi_f$ & Flow parameter\\
$Re$ & Reynolds number\\
$Pr$ & Prandtl number\\
$\bm{\alpha}$ & output of the branch network\\
$\bm{\phi}$ & output of the trunk network\\
$\tau_w$ & Wall Shear Stress\\
$L$ & Lift\\
$D$ & Drag\\
$\protect \overrightarrow{n}$ & Unit Normal\\
$\protect \overrightarrow{t}$ & Unit Tangent \\
\end{longtable}}

\newpage
\input{introduction}

\input{problem_setup}

\input{methodology}

\input{results}

\input{conclusion}

\section*{Acknowledgments}
The research reported in this document is performed in connection with cooperative agreement contract/instrument W911NF-22-2-0047 with the U.S. Army Research Laboratory. L.B. and A.G. were supported by the US Army Research Laboratory 6.1 basic research program in vehicle power and propulsion sciences. The views and conclusions contained in this document are those of the authors and should not be interpreted as representing the official policies or positions, either expressed or implied, of the U.S. Army Research Laboratory or the U.S. Government. The U.S. Government  is  authorized  to  reproduce  and  distribute  reprints  for  Government  purposes  notwithstanding  any copyright notation herein.

\input{appendix}

\newpage
\bibliographystyle{unsrt}  
\bibliography{reference}

\end{document}

%% file: abstract.tex
\begin{abstract}

\textmd{Deep neural operators, such as DeepONets, have changed the paradigm in high-dimensional nonlinear regression from function regression to (differential) operator regression, paving the way for significant changes in computational engineering applications. Here, we investigate the use of DeepONets to infer flow fields around unseen airfoils with the aim of shape optimization, an important design problem in aerodynamics that typically taxes computational resources heavily.  
We present results which display little to no degradation in prediction accuracy, while reducing the online optimization cost by orders of magnitude. We consider NACA airfoils as a test case for our proposed approach, as their shape can be easily defined by the four-digit parametrization. We successfully optimize the constrained NACA four-digit problem with respect to maximizing the lift-to-drag ratio and validate all results by comparing them to a high-order CFD solver. We find that DeepONets have low generalization error, making them ideal for generating solutions of unseen shapes. Specifically, pressure, density, and velocity fields are accurately inferred at a fraction of a second, hence enabling the use of general objective functions beyond the maximization of the lift-to-drag ratio considered in the current work. 
} 

\end{abstract}

%% file: introduction.tex
\section{Introduction}

\begin{figure}[H]
\centering
    \includegraphics[scale=0.6]{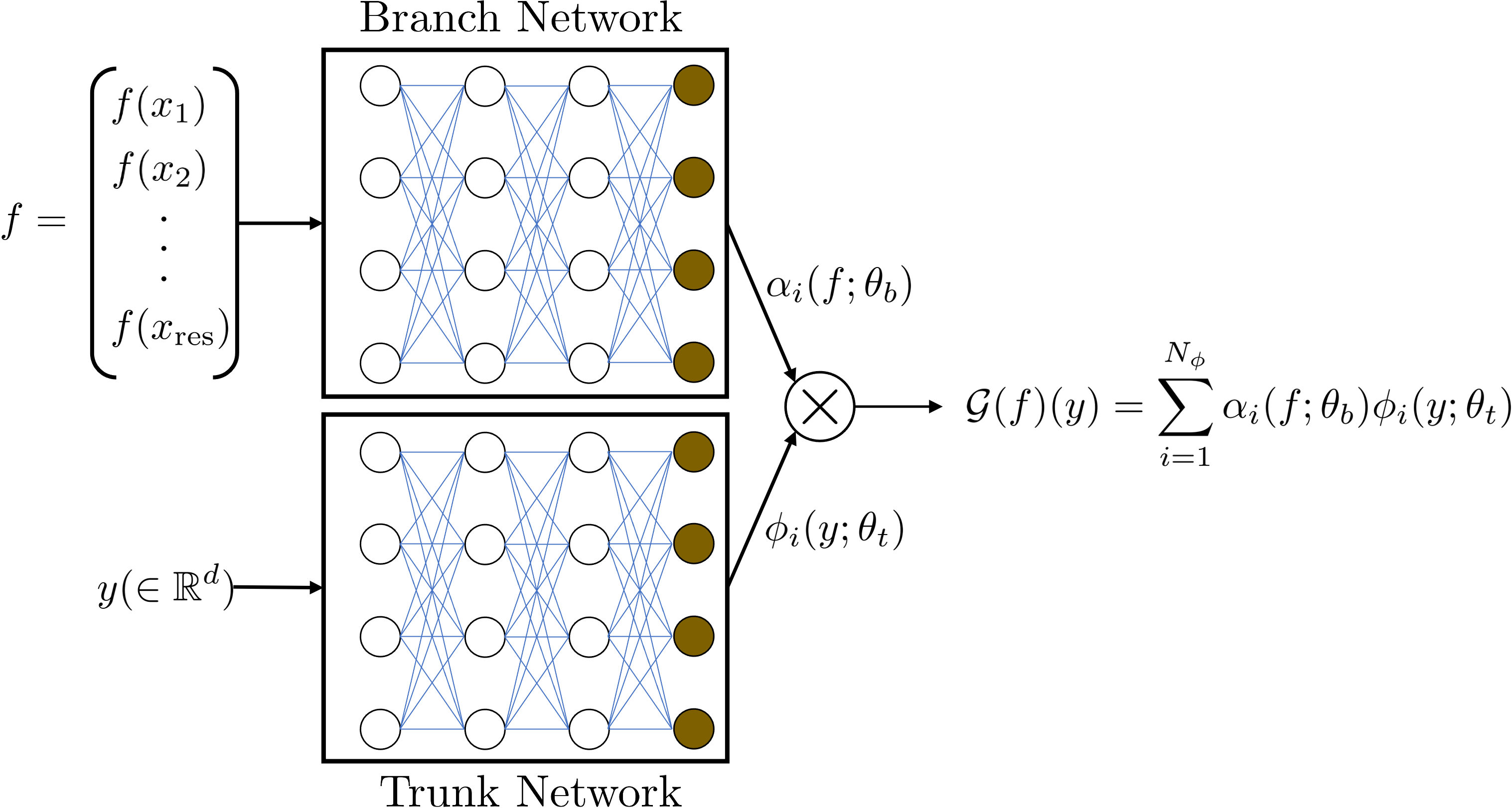}
  \caption{\textbf{Schematic of a DeepONet.}\textmd{ $f$ represents a general input function provided to the DeepONet. A discrete representation of $f$ with resolution (=res) is provided as the input to the branch network. In this figure, $y(\in \mathbb{R}^d)$ represents the domain coordinates of the output function.}} 
  \label{fig: DeepONet_diagram}
\end{figure}

Neural networks that solve regression problems map input data to output data, whereas neural operators map functions to functions. This recent paradigm shift in perspective, starting with the original paper on the deep operator network or DeepONet \cite{lu2021learning,lu2019deeponet}, provides a new modeling capability that is very useful in engineering -- that is, the ability to replace very complex and computational resource-taxing multiphysics systems with neural operators that can provide functional outputs in real-time. Specifically, unlike other physics-informed neural networks (PINNs) \cite{raissi2019physics} that require optimization during training and testing, a DeepONet does not require any optimization during inference, hence it can be used in real-time forecasting, including design, autonomy, control, etc. An architectural diagram of a DeepONet with the commonly used nomenclature for its components is shown in Figure \ref{fig: DeepONet_diagram}.
DeepONets can take a multi-fidelity or multi-modal input \cite{de2022bi, howard2022multifidelity, lu2022multifidelity, jin2022mionet, zhu2022reliable} in the branch network and can use an independent network as the trunk, a network that represents the output space, e.g. in space-time coordinates or in parametric space in a continuous fashion. In some sense, DeepONets can be used as surrogates in a similar fashion as reduced order models (ROMs) \cite{hesthaven2018non, hesthaven2016certified, benner2017model, williams2015data, chiavazzo2014reduced, lieberman2010parameter, bui2008model, benner2015survey, amsallem2015design, carlberg2008compact, choi2020gradient}. However, unlike ROMs, they are over-parametrized which leads to both generalizability and robustness to noise that is not possible with ROMs, see the recent work of \cite{kontolati2022influence}. 

In the present work, we investigate the possibility of using DeepONets for representing functions over different solutions domains, something that has not been explored before. In particular, we focus on aerodynamic design by  considering a classical problem of optimizing the shape of an airfoil at subsonic flow conditions.
Aerodynamic shape optimization (ASO) of airfoils plays a vital role in the design of efficient modern commercial aircraft. The aerodynamic geometric optimization process is usually applied to the airfoil shape that forms the cross sections of the three-dimensional (3D) airfoil. The geometric optimization is performed to reduce the drag force while increasing the lift to enhance the aircraft's fuel efficiency, hence reducing the transportation cost. The optimization process often requires a numerical model that predicts the flow field around a given airfoil geometry and computes the lift and drag for desired flow conditions. Traditionally, the numerical models are compressible flow numerical solvers, which are computationally intensive to accurately realize the flow field around a complex airfoil. Surrogate models can be introduced to circumvent the time-consuming part of the optimization loop where the numerical solver calculates aerodynamic forces. 

ASO traditionally uses two main paradigms, gradient-based and gradient-free optimization approaches. The gradient-based approaches require the calculation of the cost function derivative with respect to the design variables. When the number of design variables exceeds a certain threshold, the gradient-based optimization becomes infeasible due to its expensive computational cost \cite{yu2018influence}. The adjoint formulation is derived from either the Euler or Navier-Stokes equations to make the gradient optimization independent of the design variables \cite{reuther1996aerodynamic,carpentieri2007adjoint,nadarajah2001studies,srinath2010adjoint}. Solving the adjoint equations can be as time-consuming as solving the governing equations (i.e., the rule-of-thumb is that forward and adjoint solutions together is at least twice the cost of the forward solver). Also, the optimization process can fall into local minima leading to a non-optimized geometry \cite{chernukhin2013multimodality}. Gradient-free approaches \cite{li2019surrogate,wu2022aerodynamic} can avoid local minima by employing the direct optimization approach that uses many costly numerical simulations of the compressible flows. The numerous simulations of the flow can help achieve the global minimum of the cost function but at the expense of high computational costs. Surrogate models can be deployed to realize the flow field with acceptable accuracy and an immense speedup compared to the full CFD simulations. The surrogate model can then be used in both a gradient-based of a gradient-free global optimization process. The surrogate-based models are usually coupled with gradient-free optimizations such as genetic algorithm and particle swarms optimization (PSO) \cite{eberhart1995new} methods. Krige \cite{krige1951statistical} proposed the Kriging surrogate model that is employed in aerospace design experiments \cite{liu2017efficient,li2019surrogate}. The Kriging surrogate model must be trained both before and during the optimization process since the surrogate model is usually inaccurate based solely on the initial training. High generalization error of the Kriging surrogate model results in inaccurate model prediction by the initial training stage. A deep operator network (DeepONet), as proposed in this work, can alleviate this issue since it maps a function to another function, which improves the generalization error significantly \cite{lu2021learning}. 

The parameterization of the geometry drastically affects ASO's computational cost and accuracy. Parameterizing the geometry reduces the number of design variables through which the optimization algorithm must search. Reducing the number of design variables simplifies the optimization process and decreases the sensitivity to noise. The parametric model must reproduce a wide range of airfoil shapes and keep the number of design variables minimal. Various geometric parametric models have been employed for ASO. Carpentieri \emph{et al.} \cite{carpentieri2007adjoint} employed orthogonal Chebyshev polynomials to construct the airfoil curves. An orthogonal polynomial is used to cover the entire design space. Lepine \emph{et al.} \cite{lepine2001optimized} used Non-Uniform Rational Basis Spline (NURBS) to parameterize a large class of airfoil shapes by only using $13$ control points. Following the Lepine \emph{et al.} \cite{lepine2001optimized} idea, Srinath and Mittal \cite{srinath2010adjoint} also employed NURBS for the parameterization. Other researchers have used B-Spline \cite{wang2019adjoint}, and  Bezier \cite{papadimitriou2016aerodynamic} curves, Hicks and Henn's functions \cite{hicks1978wing} for airfoil shape construction. Painchaud-Ouellet \emph{et al.} \cite{painchaud2006airfoil} used NURBS for the shape optimization of an airfoil within transonic regimes. They showed that using NURBS ensures the regularity of the airfoil shape. The airfoil shape can also be constructed using a deformation method \cite{hicks1978wing}. This method adds a linear combination of bumps to a baseline airfoil shape for parameterization \cite{chen2017airfoil,he2019robust}. The Class function/shape function Transformation (CST) approach \cite{wu2019benchmark} employs Bernstein polynomials \cite{akram2021cfd} to parameterize airfoils and other aerodynamic geometries. Other researches employed proper orthogonal decomposition (POD) \cite{wu2019benchmark} to reduce the number of design variables. In the current study, we employ the NACA 4-digits airfoil and NURBS parameterizations for the airfoil shape construction.

With the significant advancement in computational power, Deep Neural Network (DNN) tools have gained much attention for developing surrogate models \cite{zhang2021multi,zhiwei2020non,renganathan2021enhanced}. The DNN approach can be readily trained for numerous input design variables to predict the cost function of the optimization loop. Du \emph{et al.} \cite{du2021rapid} trained a feed-forward DNN to receive airfoil shapes and predict drag and lift coefficients. They also used RNN models for estimating the pressure coefficient. The optimal airfoil design determined using the surrogate model was compared with an airfoil design obtained with a CFD-based optimization process \cite{du2021rapid}. Liao \emph{et al.} \cite{liao2021multi} designed a surrogate model using a multi-fidelity Convolutional Neural Network (CNN) with transfer learning. This learning method transfers the information learned in a specific domain to a similar field. The low-fidelity samples are taken as the source, and the high-fidelity ones are assigned as targets. Tao and Sun \cite{tao2019application} introduced a Deep Belief Network (DBN) to be trained with low-fidelity data. The trained DBN was later combined with high-fidelity data using regression to create a surrogate model for shape optimization. Existing surrogate models for shape optimization are all trained to predict lift, drag, or pressure coefficients. In contrast, the flow field around the aerodynamic shape is not inferred. Here, we construct a surrogate model that predicts the flow field around the airfoil shape using a DeepONet. Predicting the flow field provides additional information that can be used in the cost function of the optimization loop.
We aim to develop an aerodynamic shape optimization framework using a surrogate model that can infer the flow field around the geometry. The surrogate model is constructed using a DeepONet and is trained using high-fidelity CFD simulations of airfoils in a subsonic flow regime. The surrogate model is then implemented in two different optimization frameworks for shape optimization. The novelties of this study include the following:

\begin{itemize}
     \item Generating a DeepONet-based surrogate model, which is an efficient and inexpensive instantiation of the exorbitant CFD solver.
    \item The construction of the surrogate model is invariant to input space, which can be defined as low or high-dimensional parameterizations.
    \item Prediction of high-dimensional flow field can be used for various cost functions in the optimization loop.
    \item Drag and lift coefficients are computed using the inferred high-dimensional flow field, resulting in more accurate predictions.
    \item Integration of the Dakota optimization framework with the DeepOnet surrogate.
\end{itemize}

The rest of the article is organized as follows. We begin by defining the optimization process. We then present the data generation for training the surrogate model using the open-source spectral/$hp$ element Nektar++ CFD solver. The training procedure of the DeepONet-based surrogate model is explained. Later, the optimization results using two different methods are represented. Finally, we summarize our findings in the Conclusions section.
In the Appendix, we provide verification on the accuracy of the data generated by repeating selected simulations using different codes. Additionally, we provide validation of the Dakota optimizer by comparing against multiple approachs, and find that all the approaches studied converge to the same solution.

%% file: problem_setup.tex
\section{Problem setup}
\begin{figure}
\centering
\includegraphics[scale=0.6]{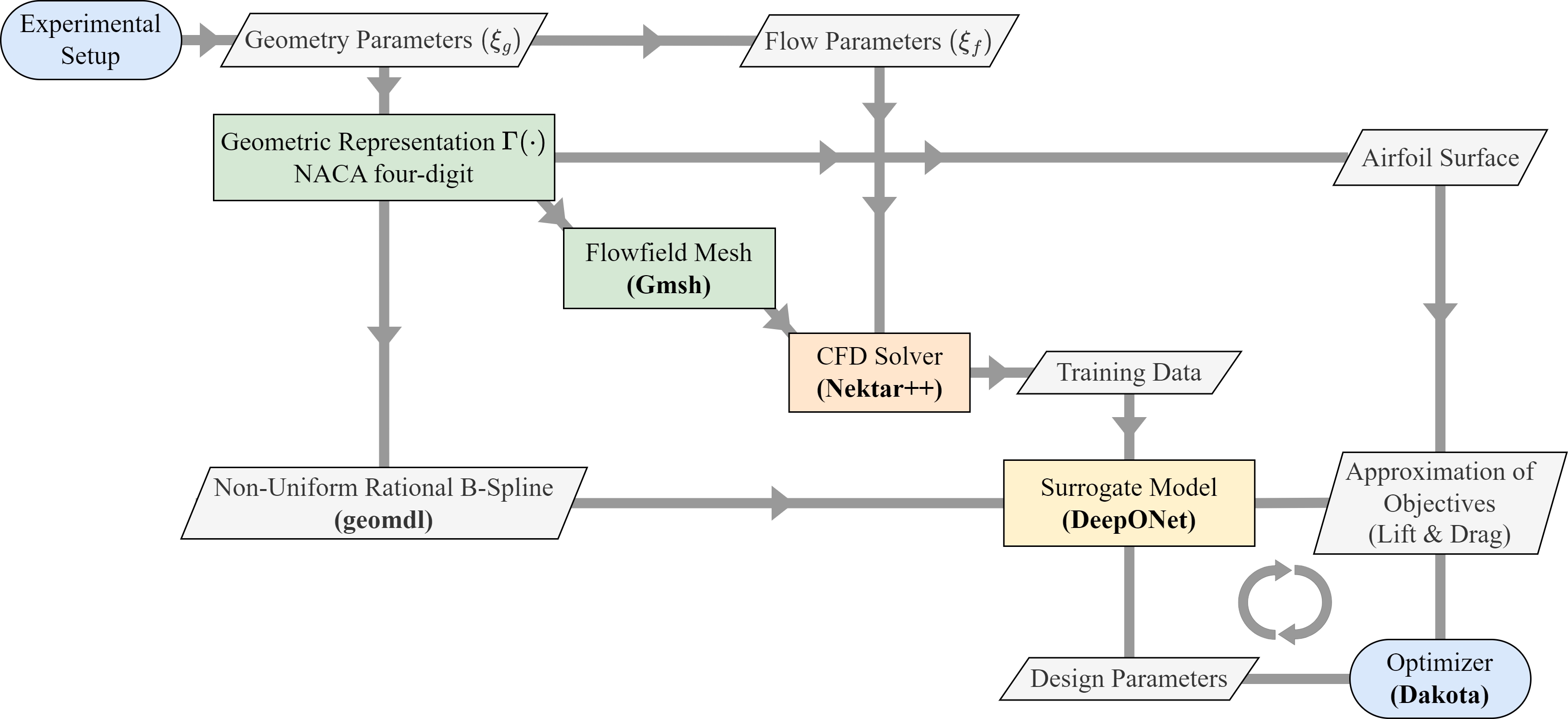}
  \caption{Diagram of airfoil geometry optimization with a DeepONet surrogate model. \textmd{The blue ovals indicate the beginning and end states of the method. The main intermediate steps are highlighted in colored boxes, whereas the auxiliary steps are grey parallelograms.}} 
  \label{fig:pipeline}
\end{figure}

To highlight the capabilities of DeepONets as function-to-function maps that can be used within the airfoil shape optimization process, we start by reviewing the traditional end-to-end shape optimization pipeline augmented with DeepONet training.  A schematic of the pipeline is shown in Figure \ref{fig:pipeline}. Reviewing the figure from upper left to lower right, we start with an experimental setup.  This represents the determination of the feasible set from which the parametric airfoils in training will be drawn, the aerodynamic conditions, and any other engineering constraints related to the problem. We choose NACA four-digit airfoils as our geometric representation, which provides the upper and lower surface equations, given a random draw of parameters. This representation is then used in three places: (1) to directly mesh the flowfield around the airfoil for which we use Gmsh; (2) to use in querying the surrogate model on the surface of the airfoil when predicting the objective lift-to-drag ratio; and (3) as the branch input to the DeepONet function-to-function map, which is pre-processed using NURBS to lower the dimensionality. 

In terms of creating the surrogate model, this can be viewed as the following forward problem. One deciphers the geometric and flow parameters, which are then used to create the inputs to a CFD solver: a geometric representation of the airfoil and its corresponding mesh to be used for approximating the flowfield and a flow parameter file. These are then input to a flow solver -- in our case, the CFD solver Nektar++. Results from this solver are used to generate training data used to train our DeepONet surrogate.  Finally, the lower-right quadrant of the diagram denotes the shape optimization process using the trained DeepONet surrogate. This process is recursive as the optimizer, for which we use Dakota, queries the DeepOnet surrogate model with new design parameters until the objective function is sufficiently minimized. This process differs from existing approaches because the bulk of the optimization process is done offline. Generating data using the CFD solver can be expensive, but with the final trained DeepOnet, the online cost of geometric optimization is orders of magnitude faster than other methods. Furthermore, as long as the objective can be created by the DeepONet flowfields trained over, a different objective function can be defined, not only lift-to-drag, and an airfoil can be quickly optimized with respect to the new objective without any additional cost.

For the purposes of our experiments, we have focused on using 2D compressible Navier-Stokes fields at Reynolds number $Re = 500$ and Mach number $Ma = 0.5$ for our training.  These values have been chosen to allow us to focus on DeepONet's ability to capture variations in domains (instead of the compounding effects of unsteadiness, etc.). Given the success of DeepONets under this experimental setup, future work will extend this pipeline to more complex flows and experimental conditions, such as varying the angle of attack, morphing geometry, handling unsteady flow, or going into the high speed flow regimes.

%% file: methodology.tex
\section{Methodology}
\subsection{Data Generation}


For each example in the dataset, we define a set of geometric parameters $(\xi_g)$ and flow parameters $(\xi_f)$. The geometric parameters are then converted into another representation, such as surface coordinates derived from the NACA airfoil equations; this transformation is given by $\Gamma(\xi_g)$. These points are then used to mesh the flowfield domain with Gmsh \cite{gmsh}, which is then input into the flowfield simulation software Nektar++ \cite{CANTWELL2015205, MOXEY2020107110}. We obtain the solutions fields from Nektar++ through post-processing for density, x-velocity, y-velocity, and pressure. This is saved in two sets, one in a subdomain around the airfoil for training the DeepONet, and one at airfoil sensors on the surface for validation. We also save the Nektar++ lift and drag forces for validation of the discrete integration of the airfoil forces. 


\subsubsection{Geometry generation}
\begin{figure}
\centering
\includegraphics[scale=0.4]{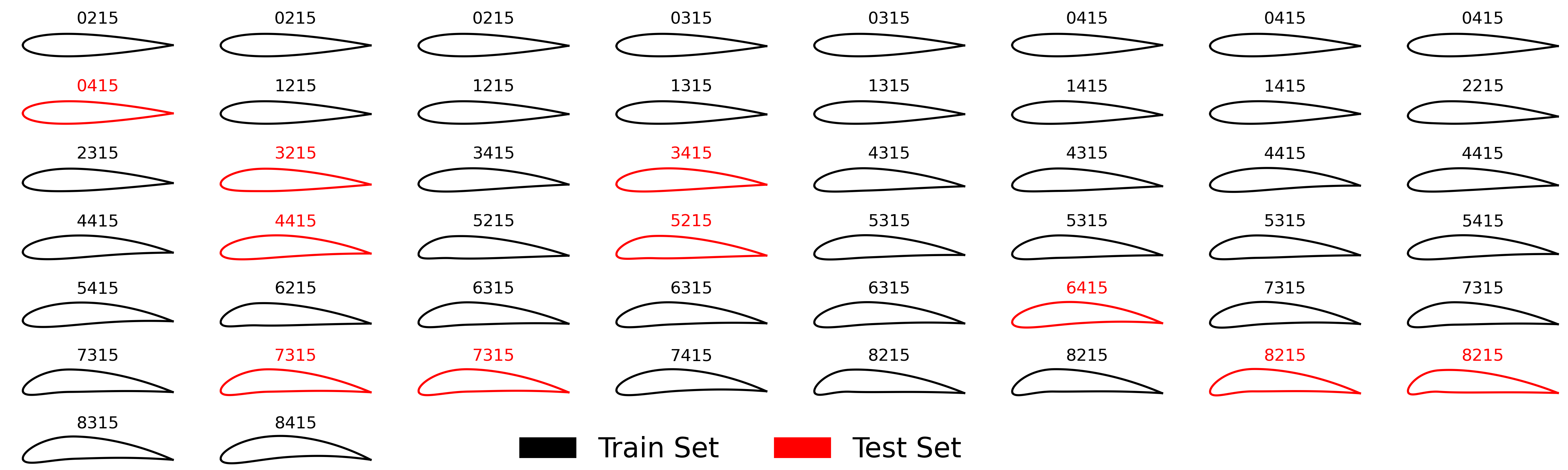}
  \caption{Train and test set geometries sampled over the $\xi_g$ domain. \textmd{Note that the numbers correspond to NACA airfoils and duplicate numbers are due to rounding to the nearest integer for readability. The true NACA parameters are drawn from a uniform distribution and are, therefore, real-valued.}}
  \label{fig:dataset}
\end{figure}

NACA 4-digit airfoils provide an excellent testbed application for geometry optimization using DeepONets since they can represent a wide range of shapes from well-known and studied parameterized geometric equations. Our geometry optimization framework could be easily extended to any parameterized geometry, such as NACA 5-digits or beyond. Following \cite{jacobs_ward_pinkerton_1933}, we define the parametric equations for the surface of an airfoil with a chord length of one, as follows:
\begin{align}
& y_t = \frac{t}{0.2} \left( a_0 \sqrt{x} + a_1 x + a_2 x^2 + a_3 x^3 + a_x x^4 \right) \label{eq:y_t} \\
& y_c =\begin{cases}
          \frac{m}{p^2} \left( 2px - x^2\right) \quad &\text{if} \; x < p \\
          \frac{m}{\left(1-p \right)^2} \left(1 - 2p + 2px - x^2 \right)\quad &\text{if} \; x > p \\
     \end{cases}     
& \theta = \tan^{-1} \left( \frac{dy_c}{dx} \right) \\
& x_u =  x - y_t \sin(\theta), \quad y_u =  y_c + y_t \cos(\theta) \notag \\
& x_l =  x + y_t \sin(\theta), \quad y_l =  y_c - y_t \cos(\theta) \label{eq:xy_l}
\end{align}
where $a_0 = 0.2969, a_1 = -0.1260, a_2 = -0.3516, a_3 = 0.2843, a_4 = -0.1015$. We can therefore define our geometry as a point cloud with coordinate sets $\left (x_u, y_u, x_l, y_l \right)$ parameterized by $\xi_g = \left( t, p, m \right)$. A series of $x$ locations are found using cosine spacing with $100$ points; this increases the geometric fidelity around the leading and trailing edge, increasing the accuracy of the mesh and flowfield simulation at these important locations. To simplify the problem and reduce the likelihood of flow separation or turbulence, we constrain $\xi_g$. The maximum thickness ($t$) is set to a constant $0.15$, and the domain of the parametric space left by the position of maximum camber ($p$) and maximum camber ($m$) is $p \times m \in [0.2, 0.5] \times [0.0, 0.09]$. Therefore, for one geometric example in either the train or test set, we draw a $\xi_g$ tuple where the variable parameters are drawn from a uniform distribution within their domains. 
We perform this draw $50$ times and obtain the surface coordinates from Equations \ref{eq:y_t} - \ref{eq:xy_l}, splitting it into $40$ training and $10$ testing examples as seen in Fig. \ref{fig:dataset}. The test/train split is an essential aspect of deep learning; here, we choose a relatively sparse sampling highlighting the ability of DeepONets to generalize well to unseen parameters. 

Next, to lower the input dimensionality into the DeepONet branch, we fit the airfoil surface with Non-Uniform Rational B-Splines (NURBS) with $30$ control points using geomdl \cite{bingol2019geomdl}. This reduces the input dimensionality from $200$ $(x,y)$ pairs to only $30$. 



\subsubsection{Mesh generation}
\begin{figure}
\centering
\includegraphics[scale=0.4]{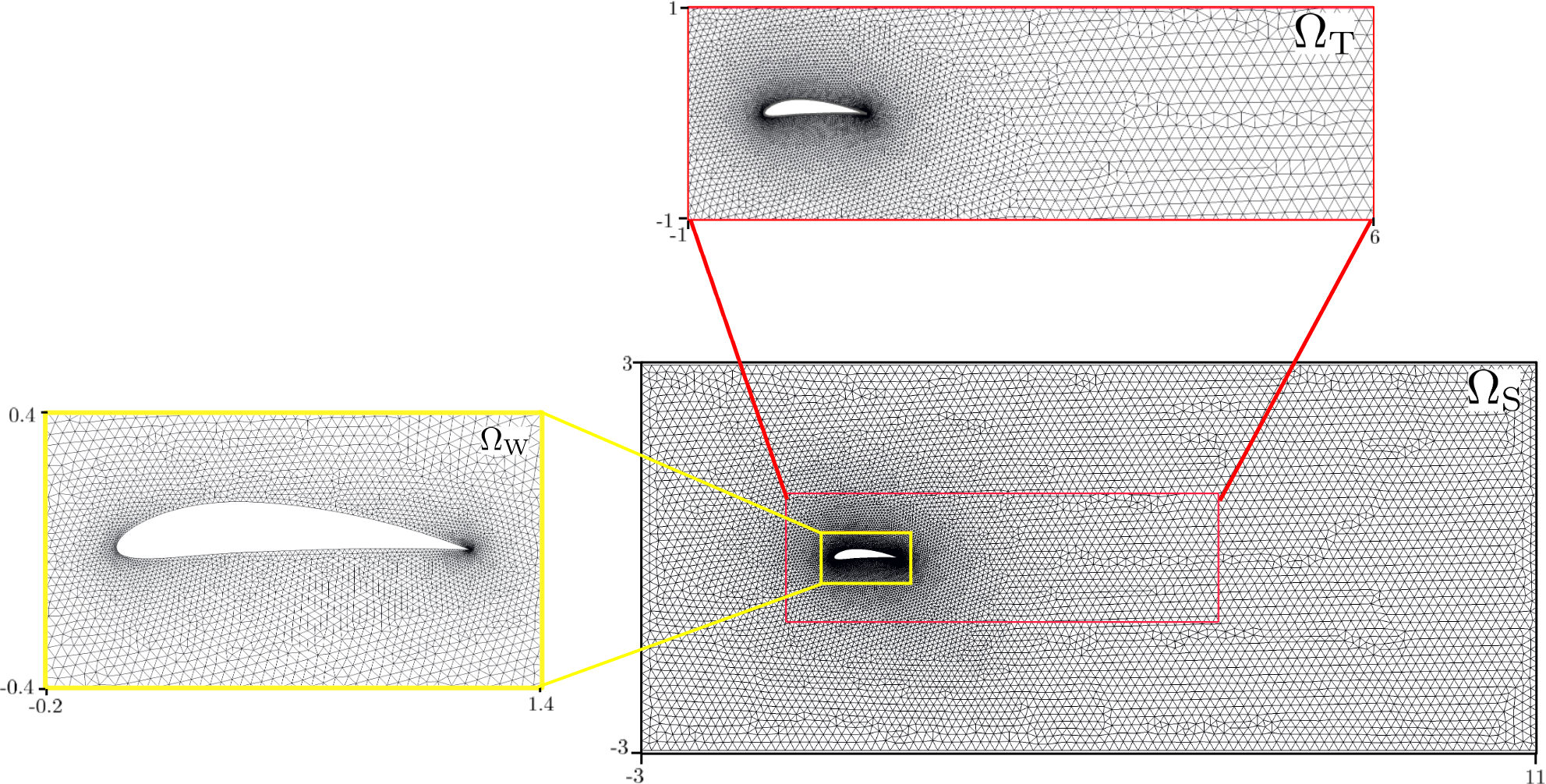}
  \caption{Discretization of an airfoil with bounding domain. \textmd{$\Omega_{\rm S}$ represents the entire domain over which the compressible Navier-Stokes equations are  solved. $\Omega_{\rm T}$ (inset image with red border) represents the domain for which a DeepONet is trained. Inset image with yellow boundary represents the subdomain  $(\Omega_{\rm W})$ showing the mesh refinement around the airfoil.}}
  \label{fig:mesh}
\end{figure}

The meshes are generated using Gmsh \cite{gmsh} for parametrized airfoils geometry with a minimum characteristic length of $0.01$ at parametrized locations. The airfoil flowfields are meshed using the $200$ surface points exactly from the NACA equations, not the NURBS fit, which contains some inaccuracy. A spline function is used to represent the 1D geometry of the boundaries of airfoils. To resolve the flow at leading and trailing edges, meshes are refined by using the splitting approach \cite{mark2008computational}. The NURBS low-dimensional representation is a step to reduce overparameterization in the DeepONet, which is unnecessary for generating high-fidelity training data. The mesh generation for all $50$ airfoils is automated using Gmsh's Python API integrated with the geometry generation in Python, so no manual operations are needed. Fig. \ref{fig:mesh} shows the mesh of the entire simulated domain $\Omega_S$, which is then input into Nektar++ along with the flow parameters to generate the DeepONet training data. As seen in the figure, only a subset of the solved steady-state domain $\Omega_T$ is used in training the DeepONet. This is because the DeepOnet is a function-to-function map and does not strictly obey boundary conditions or is affected by phenomena such as reflections due to the boundaries. It is performing regression on the dataset, not solving the system of equations, and therefore can be taken as a smaller domain, even without freestream conditions. Since the objective is geometric optimization, this subdomain simplifies the DeepONet training problem and cost of training.

\subsubsection{Flowfield simulation}
We used a compressible flow solver implemented in Nektar++ to generate the flow field data.
The Nektar++ (www.nektar.info) is an open-source software
framework \cite{CANTWELL2015205, MOXEY2020107110} designed to support the development of high-performance, scalable solvers for partial
differential equations using the spectral/hp element method.  The 2D compressible flow solver uses the 
two-dimensional compressible Navier-Stokes equations expressed as,
\begin{align} \label{eq-system}
\begin{aligned}
&\frac{\partial {\rho}}{\partial {t}} + \frac{\partial {\rho} {u}}{\partial {x}} +  \frac{\partial {\rho} {v}}{\partial {y}} = 0 \\
&\frac{\partial {\rho} {u}}{\partial {t}} + \frac{\partial {\rho} {u}^{2} + {p}}{\partial {x}} +  \frac{\partial {\rho} {u}{v}}{\partial {y}} 
= \frac{1}{Re} \left( \frac{\partial {\tau}_{xx}}{\partial {x}} + \frac{\partial {\tau}_{yx}}{\partial {y}} \right)\\
&\frac{\partial {\rho} {v}}{\partial {t}} + \frac{\partial {\rho} {u}{v}}{\partial {x}} +  \frac{\partial {\rho} {v}^{2} + {p}}{\partial {y}} 
= \frac{1}{Re} \left( \frac{\partial {\tau}_{xy}}{\partial {x}} + \frac{\partial {\tau}_{yy}}{\partial {y}} \right)\\
&\frac{\partial {E}}{\partial {t}} + \frac{\partial ({E}+{p}){u}}{\partial {x}} +  \frac{\partial ({E}+{p}){v}}{\partial {y}} 
= \frac{1}{Re} \left[ \frac{\partial \left( {u}{{\tau}}_{xx}+{v}{{\tau}}_{xy}+ {{\kappa}}\tfrac{\partial {T}}{\partial {x}} \right)}{\partial {x}} 
+ \frac{\partial \left( {u}{{\tau}}_{xy}+ {v}{{\tau}}_{yy} + {{\kappa}}\tfrac{\partial {T}}{\partial {y}} \right)}{\partial {y}} 
\right],
\end{aligned}
\end{align}
where ${p}={R}{\rho}{T}$, ${R} = \frac{1}{\gamma Ma^2}$, ${k} = \frac{\gamma}{\gamma-1} \frac{{{\mu}} {R}}{Pr}$, with ${{\mu}}$ being the non-dimensional viscosity and computed by using the Sutherland law as 
\begin{align}
{{\mu}} = \frac{{T}^{3/2}}{Re} \frac{1 + C/T_{\infty}}{{T} + C/T_{\infty}}. \label{eq:mu}
\end{align}
Expressions for $\tau_{xx}$, $\tau_{yy}$, and $\tau_{xy}$ are as follows
\begin{align*}
\tau_{xx} &= 2\mu \left(u_{x} - \frac{u_{x} + v_{y}}{3} \right),\\
\tau_{yy} &= 2\mu \left(v_{y} - \frac{{u}_{x} + {v}_{y}}{3} \right),\\
\tau_{xy} &= \mu \left(u_{y} + {v}_{x} \right).
\end{align*}

We aim to simulate the flow past airfoils by solving the compressible Navier-Stokes
equations given by Equation \eqref{eq-system} with free-stream parameters $M_{\infty}=0.5$, $Re_{L=1}=500, ~u_{\infty}=1,~v_{\infty}=0,~T_{\infty}=1$, ${\rm AoA=0}$ and $Pr=0.72$. The flow domain is $[-3, 11] \times [-3, 3]$ and discretized by conforming triangular elements. To solve \eqref{eq-system}, we use the discontinuous Galerkin spectral element method (DGSEM) with basis functions spanned in 2D by Legendre polynomials of the second degree. For advection and diffusion terms, weak and interior penalty-based dG approach with Roe upwinding is used in space. A diagonally implicit Runge–Kutta (DIRK) method  is used as a time integrator for advection and diffusion terms. The boundary conditions of inflow, outflow, adiabatic wall at airfoil surface, and high-order boundary conditions at the top and bottom are imposed weekly. For detailed descriptions of solvers and methods, readers are advised to read the article by \cite{mengaldo2014guide}.
\subsection{Surrogate model: DeepONet}
\subsubsection{Brief review of DeepONets}
Neural operators are neural network models developed on the basis of the universal operator approximation theorem \cite{chen1995universal}.  The neural operators  learn the mapping between spaces of function and directly learn the underlying operator from the available training data. DeepONets \cite{lu2021learning} and Fourier Neural Operators (FNO) \cite{li2020neural} are the two popular neural operators extensively used for solving a wide spectrum of problems in diverse scientific areas. A DeepONet consists of a branch network that encodes the input function and a trunk network that learns a collection of basis functions. The DeepONet output is computed by taking the inner product between the branch and trunk network outputs. 
\subsubsection{Training and Testing of DeepONets}
We train four different DeepONet models to learn the pressure ($p$), density ($\rho$), and velocity ($u$, $v$) fields for a given airfoil geometry ($\xi_g$) from the training data. The trunk network learns a collection of basis ($\bm{\phi}$) as functions of spatial coordinates, and the branch network learns the corresponding coefficients ($\bm{\alpha}$) as a function of the airfoil geometry. The DeepONet output is defined as
\begin{equation}
    \mathcal{G}^{q}(\xi_g)(x,y) = \sum_{i=1}^{N_{\phi}} \alpha_i(\xi_g; \theta_b^q) \phi_i(x,y; \theta_t^q) \quad  q \in \{ p, \rho, u, v\}.
\end{equation}
In the case of airfoils, the geometry can be fed into the branch network by directly providing the geometric parameter, $\xi_g$. However, $\xi_g$ need not exist explicitly for a general arbitrary geometry. Under such a scenario, the geometry is often represented using the NURBS control points. To demonstrate the effectiveness of using a DeepONet in either of the situations, we investigate parameter-DeepONet that directly takes $\xi_g$ as the branch network input and NURBS-DeepONet that takes NURBS control points of the airfoil geometry as the input to the branch network. The hyperparameters of the DeepONet used in this study are provided in Table \ref{table:hyperp}.
\begin{table}  
\centering
\caption{Hyperparameters of NURBS-DeepONet and the Parameter DeepONet}
\begin{tabular}{|ll|l|ll|}
\cline{1-2} \cline{4-5}
\multicolumn{2}{|c|}{\textbf{NURBS-DeepONet}}                              &  & \multicolumn{2}{c|}{\textbf{Parameter-DeepONet}}                       \\ \cline{1-2} \cline{4-5} 
\multicolumn{1}{|l|}{Branch network   architecture:} & {[}30,100,100,50{]} &  & \multicolumn{1}{l|}{Branch network architecture:} & {[}2,100,100,50{]} \\ \cline{1-2} \cline{4-5} 
\multicolumn{1}{|l|}{Branch network   activation:}   & tanh                &  & \multicolumn{1}{l|}{Branch network activation:}   & tanh               \\ \cline{1-2} \cline{4-5} 
\multicolumn{1}{|l|}{Trunk network   architecture:}  & {[}2,100,100,50{]}  &  & \multicolumn{1}{l|}{Trunk network architecture:}  & {[}2,100,100,50{]} \\ \cline{1-2} \cline{4-5} 
\multicolumn{1}{|l|}{Trunk network   activation:}    & tanh                &  & \multicolumn{1}{l|}{Trunk network activation:}    & tanh               \\ \cline{1-2} \cline{4-5} 
\multicolumn{1}{|l|}{$N_{\phi}$:}                   & 50                  &  & \multicolumn{1}{l|}{$N_{\phi}$:}                 & 50                 \\ \cline{1-2} \cline{4-5} 
\multicolumn{1}{|l|}{Optimizer:}                     & Adam                &  & \multicolumn{1}{l|}{Optimizer:}                   & Adam               \\ \cline{1-2} \cline{4-5} 
\multicolumn{1}{|l|}{Learning rate:}               & 1.00E-04            &  & \multicolumn{1}{l|}{Learning rate:}               & 1.00E-04           \\ \cline{1-2} \cline{4-5}
\end{tabular}
\label{table:hyperp}
\end{table}
\subsubsection{Lift and Drag calculation}
We present two ways to evaluate the objective function -- in our case, lift-to-drag. The map created by the trained DeepONets takes an input geometry and spatial $(x,y)$ location in the output space and returns the value of the field at that location. So it follows that to estimate the lift and drag, we simply evaluate the discrete integral over the surface of the airfoil, for which the $(x,y)$ points are easily generated for objective function queries by Equations \ref{eq:y_t}-\ref{eq:xy_l}. The discrete integrals for lift and drag  given by Equations \ref{eq:lift} \& \ref{eq:drag} subject to the approximation of wall shear stress in Equation \ref{eq:wss}. 
\begin{align}
& \tau_w = \mu \frac{dU}{d\protect \overrightarrow{n}} \label{eq:wss} \\
& L = \int dF_x = \sum p \protect \overrightarrow{n_x} dS + \sum \tau_w \protect \overrightarrow{t_x} dS \label{eq:lift} \\
& D = \int dF_y = \sum p \protect \overrightarrow{n_y} dS + \sum \tau_w \protect \overrightarrow{t_y} dS \label{eq:drag} 
\end{align}

\begin{figure}[!ht]
\centering
\includegraphics[scale=0.4]{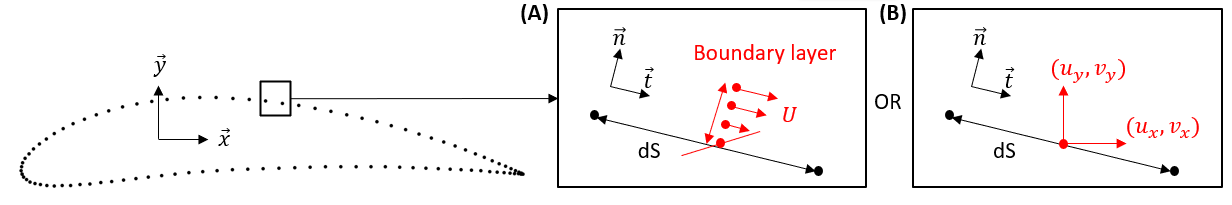}
  \caption{Illustration of the discrete integral for lift and drag. \textmd{The points in red indicate the surrogate model samples used in the construction of the approximation to $\frac{dU}{d\protect \overrightarrow{n}}$ with a finite difference approximation of the gradient or the automatic differentiation approximation using the direct network gradients.}} 
  \label{fig:integral}
\end{figure}

For the first term in the aerodynamics forces, the pressure is directly obtained from one of the DeepONet predictions. In the second term, the viscosity $\mu(p,\rho)$ is obtained by Equation \ref{eq:mu} as a function of the pressure and density DeepONets. Finally, the change in speed of the flow over the airfoil surface $\frac{dU}{d\protect \overrightarrow{n}}$, is obtained in two ways:

\begin{enumerate}
\item Finite-difference \textbf{(A)}:
\begin{align}
\frac{dU}{d \protect \overrightarrow{n}} = \frac{-U_{2} + 4U_{1} - 3U_{0}}{2h} \label{eq:fd}
\end{align}
\item Automatic-differentiation \textbf{(B)}:
\begin{align}
\frac{dU}{d\protect \overrightarrow{n}} =  (v_y - u_x) sin \theta cos \theta - v_x sin^2 \theta + u_y cos^2 \theta \label{eq:ad}
\end{align}
\end{enumerate}
where $h = 0.001$ and $\theta$ is the angle between the x-y axis and each segment's normal-tangental axis. Approach (A) is a second-order forward finite difference approximation obtained by sampling the x-velocity and y-velocity DeepONets at the appropriate locations defined by the surface normal and spacing $h$. Approach (B) utilizes the now well-known development in automatic differentiation \cite{baydin2018automatic}, primarily utilized in physics-informed machine learning for approximating partial derivatives to obtain the PDE residual. Here, since the DeepONet directly takes in the spatial $(x,y)$ coordinates and outputs the velocity components $(u,v)$, the computational graph is complete and the partials $(u_x, u_y, v_x, v_y)$ can be estimated with this method. The required sampling for each method is shown in Fig. \ref{fig:integral}. While (A) takes three times the amount of point evaluations, (B) requires the gradients to be computed, so the cost of each can be viewed as similar. However, the flexibility of using automatic differentiation in this way may allow for more complex objective functions in the future given the right mapping and subsequent computational graph. 

%% file: results.tex
\section{Results}

\subsection{Results from DeepONet model}

\begin{figure}
\centering
\includegraphics[scale=0.3]{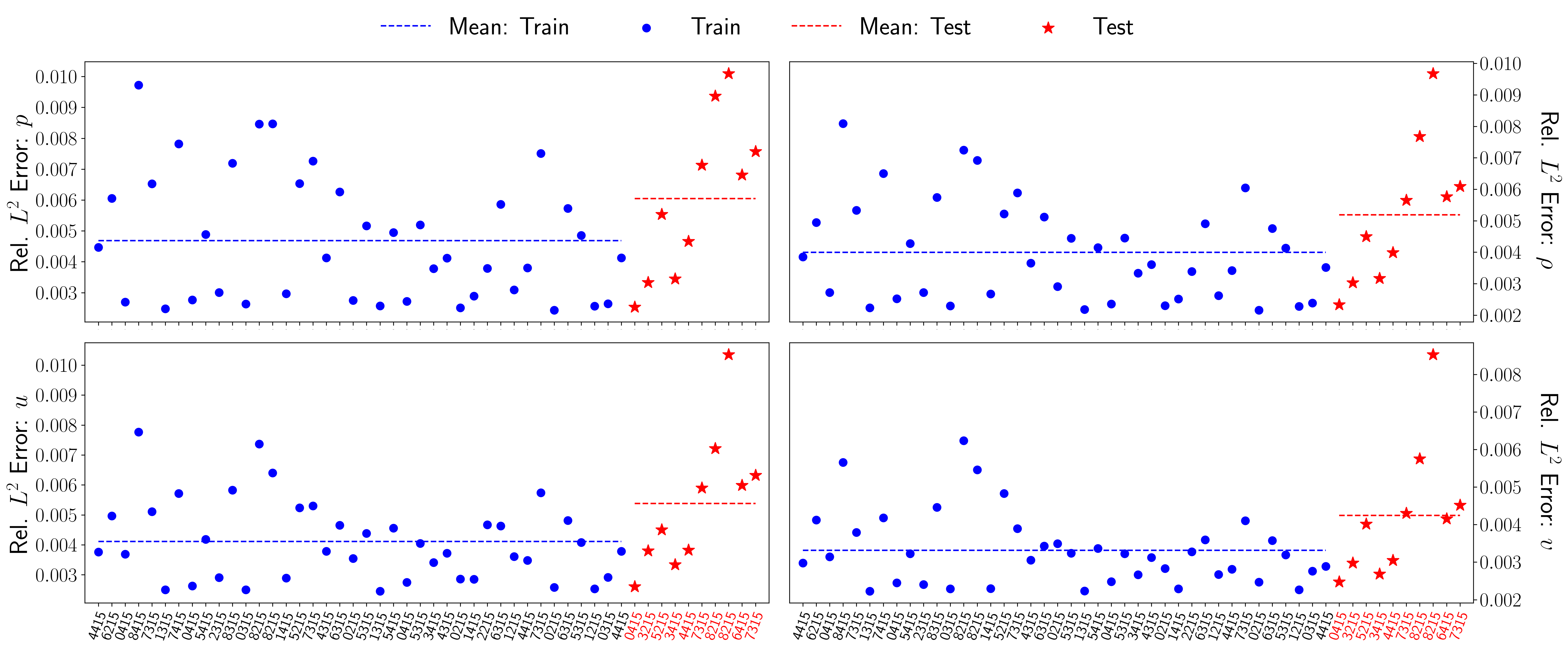}
  \caption{Error Scatter Plots. \textmd{The relative $L^2$ error corresponding to each of the train and test samples for pressure, velocity, and density fields, with respect to the DeepONet predictions, are shown in this figure.}}
  \label{fig: error_scatter_plot}
\end{figure}

\begin{figure}
\centering
\includegraphics[scale=0.7]{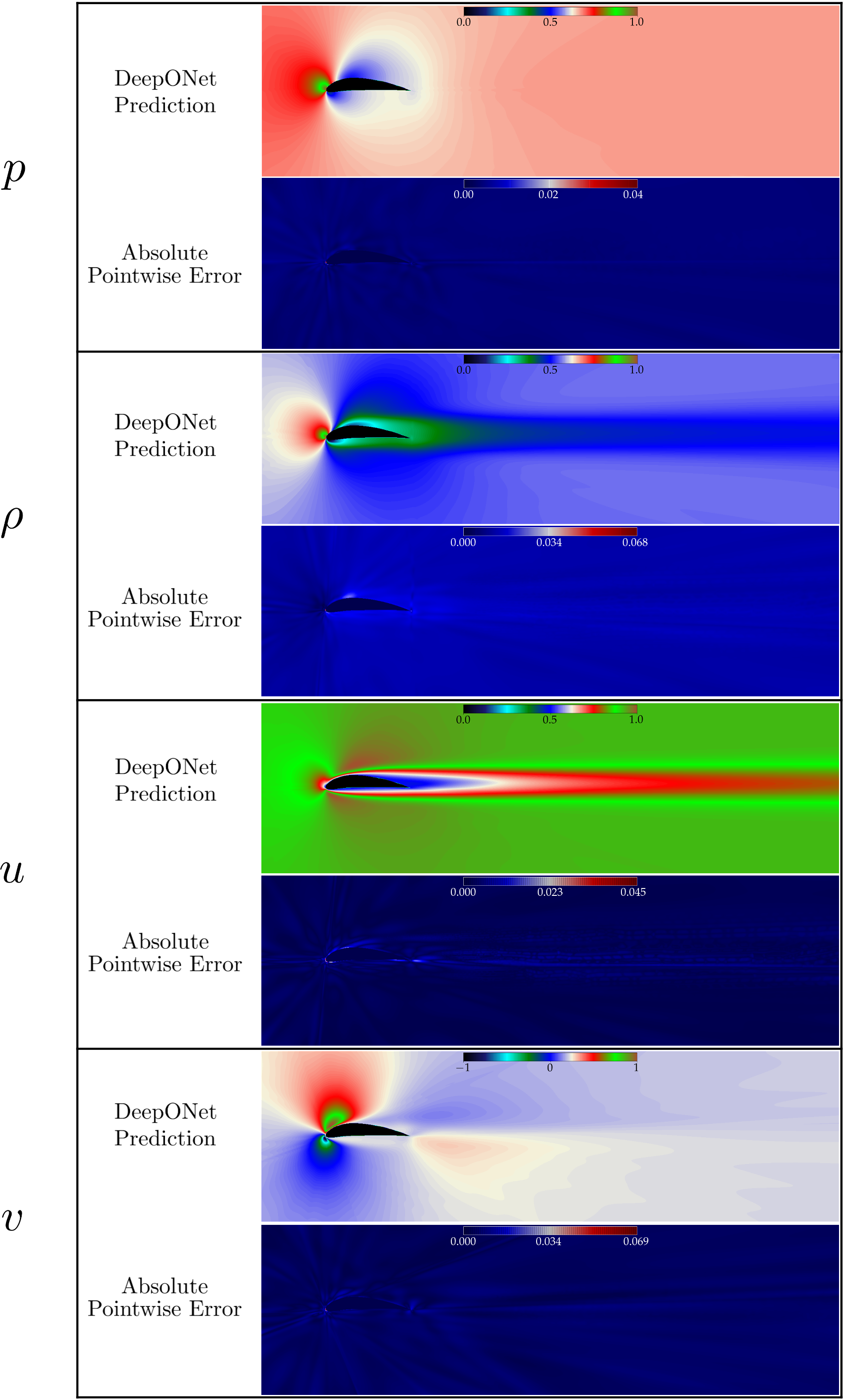}
  \caption{DeepONet Predictions. \textmd{The pressure, density, and velocity fields around the test set airfoil NACA 7315 predicted by the DeepONet, and the corresponding pointwise absolute errors are also provided.}}
  \label{fig: DeepONet_results}
\end{figure}

The training and testing relative $L^2$ error of the NURBS and parameter-based DeepONets for all four different fields are reported in table \ref{table: DeepONet results}. We observe that the NURBS and parameter-based DeepONets predict fields with similar accuracy. NURBS-DeepONet has marginally better predictions on the pressure field, while parameter-DeepONet generates marginally better predictions for velocity and density fields. The main takeaway is that either representation is sufficient for the geometry optimization of this experiment. However, we must consider that in the future more complex geometries may be used, particularly in the sense of local morphing. Therefore, the NURBS representation will likely be necessary as a direct parameter mapping may miss local nuances. The predicted fields and the absolute pointwise error by the best DeepONet models for the flowfield parameters are shown in Fig. \ref{fig: DeepONet_results}. It can be seen that the global prediction is, in general, accurate; the error is primarily localized to the airfoil's leading edge. In the future, adaptive weighting schemes will be used to improve DeepONet training, particularly at the points of difficulty, such as the surface and leading edge. The corresponding relative $L^2$ error of the fields over the entire dataset is shown in Fig. \ref{fig: error_scatter_plot}. We observe minimal generalization error and that DeepONets are globally accurate. 

\begin{table} 
\centering
\caption{Relative $L^2$ errors of the state variables trained DeepONet models}
\begin{tabular}{|l|ll|ll|}
\hline
             & \multicolumn{2}{c|}{NURBS-DeepONet}                                 & \multicolumn{2}{c|}{Parameter-DeepONet}                             \\ \hline
             & \multicolumn{1}{l|}{Train rel. $L^2$ Error} & Test rel. $L^2$ Error & \multicolumn{1}{l|}{Train rel. $L^2$ Error} & Test rel. $L^2$ Error \\ \hline
$\mathcal{G}^{p}$   & \multicolumn{1}{l|}{4.68e-03}               & 6.05e-03              & \multicolumn{1}{l|}{5.23e-03}               & 6.85e-03              \\ \hline
$\mathcal{G}^{u}$   & \multicolumn{1}{l|}{4.97e-03}               & 6.21e-03              & \multicolumn{1}{l|}{4.12e-03}               & 5.38e-03              \\ \hline
$\mathcal{G}^{v}$   & \multicolumn{1}{l|}{3.73e-03}               & 4.60e-03              & \multicolumn{1}{l|}{3.31e-03}               & 4.25e-03              \\ \hline
$\mathcal{G}^{\rho}$ & \multicolumn{1}{l|}{4.57e-03}               & 5.89e-03              & \multicolumn{1}{l|}{4.00e-03}               & 5.18e-03              \\ \hline
\end{tabular}
\label{table: DeepONet results}
\end{table}

\begin{figure} 
\centering
\includegraphics[scale=0.3]{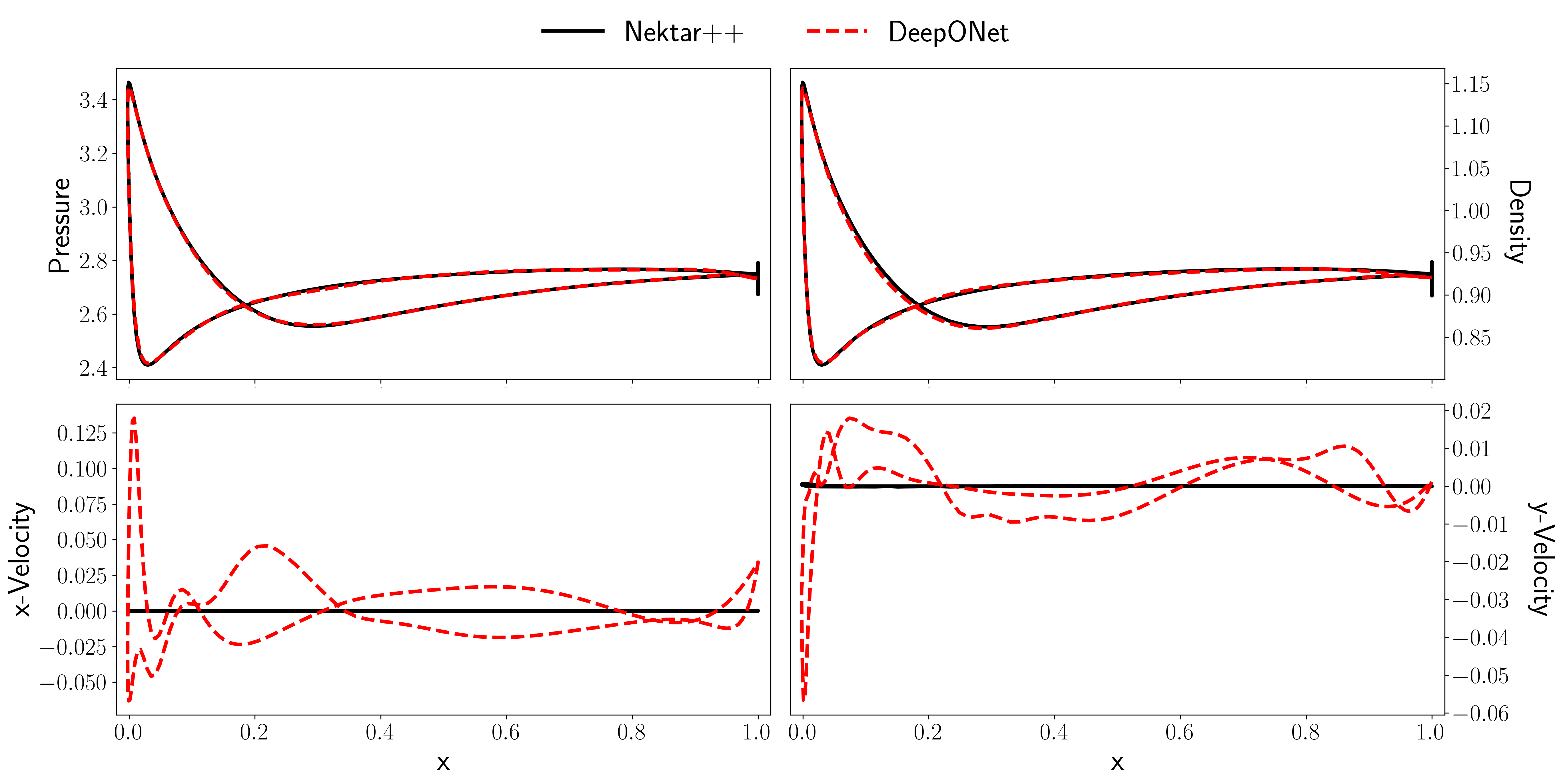}
  \caption{Plot of the flowfield variables on the surface of the test set airfoil NACA $\mathbf{7315}$. \textmd{All plots display accurate predictions on the surface, which are then used to compute the lift and drag forces. The no-slip condition is not directly enforced by the DeepONet, which results in the velocity plot difference. However, it can be seen by the y-scale that the prediction is close to zero, aside from the leading edge, and does not greatly affect the overall lift and drag computation.}}
  \label{fig:surface_prediction}
\end{figure}

\begin{figure} 
\centering
\includegraphics[scale=0.3]{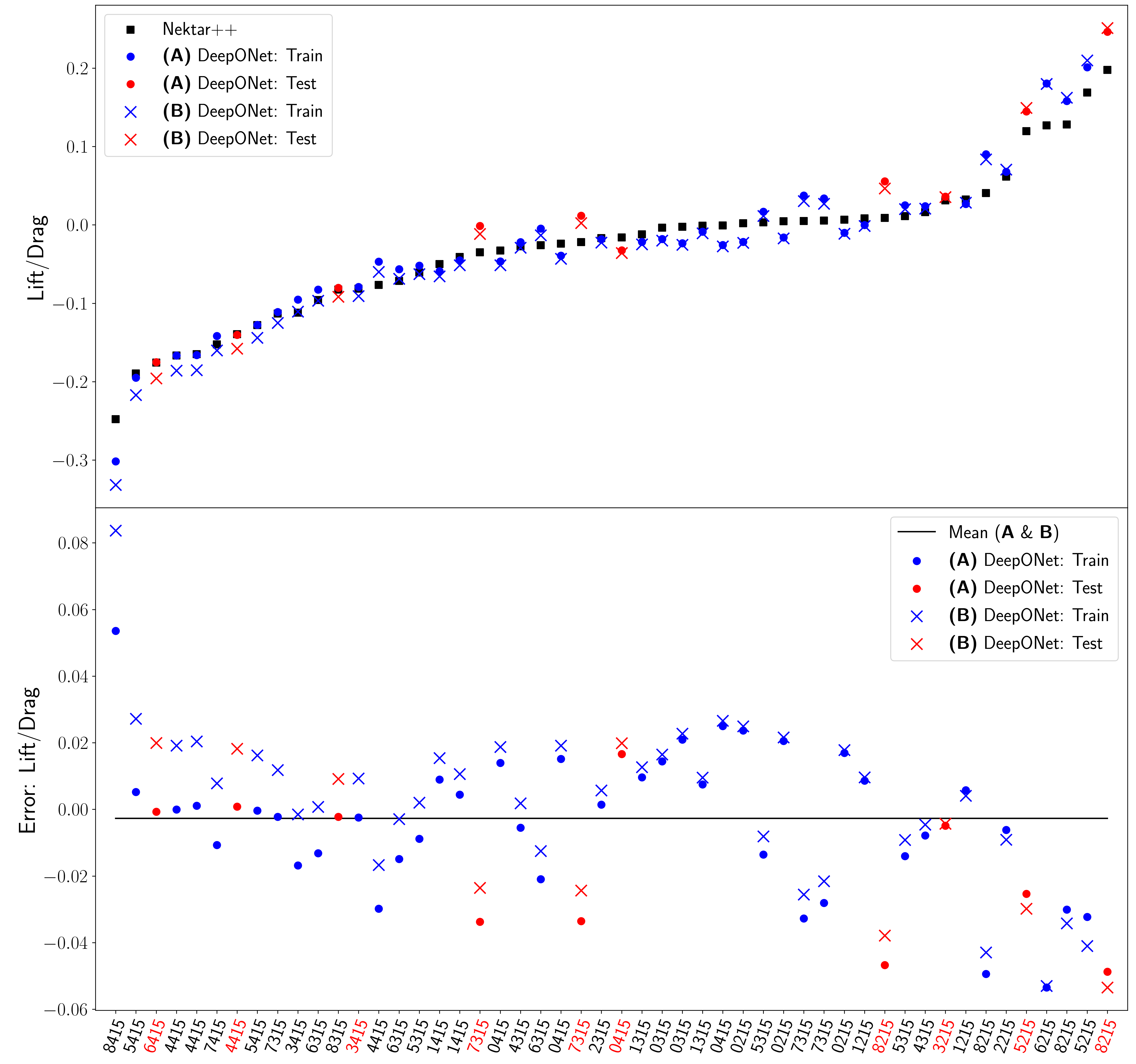}
  \caption{Plot of the computed lift-to-drag objective for the entire dataset sorted by the Nektar++ reference values. \textmd{As seen in both plots, the approximation to the high-fidelity CFD solution is very accurate and consistent throughout the entire parametric domain. Particularly, we note that the testing set performs comparably to the training set, meaning there is little to no generalization error, which is necessary when inferring unseen queried geometries during optimization.}}
  \label{fig:L2D-CFD-DON}
\end{figure}

Regarding geometry optimization, we must concern ourselves not only with the global flowfield accuracy, but also with the accuracy on the surface of the airfoil in particular. Fig. \ref{fig:surface_prediction} shows the corresponding surface prediction plots for the same airfoil presented in Fig. \ref{fig: DeepONet_results} as a function of the x-direction over the airfoil. We observe good accuracy in the pressure and density fields, which will provide very accurate predictions of the lift and drag force components due to pressure as well as the viscosity $\mu(p, \rho)$, which is a function of these fields per Equation \ref{eq:mu}. The surface's $x$ and $y$ velocity fields do not agree because the DeepONet does not strictly obey a no-slip condition. However, aside from the leading edge, the predictions are close to zero. Furthermore, the fields are not directly related to the objective lift-to-drag but indirectly related through the estimate of the change in flow speed over the surface $\frac{dU}{d\protect \overrightarrow{n}}$ obtained by approaches (A) and (B) in Equation \ref{eq:fd} and \ref{eq:ad}. Therefore, the inaccuracy does not significantly affect the overall objective prediction. This is corroborated by Fig. \ref{fig:L2D-CFD-DON}, which shows the sorted lift-to-drag ratio for the entire dataset. The results of both numerical integration approaches (A) and (B) are very accurate to the stored lift-to-drag results from the Nektar++ data generation step. We can also see in the error plot that it is quite uniform, and there are no discernible biases in the geometric parameter space, indicating we have learned the entire space well enough for the final optimized result to be accurate.  

\subsection{Shape Optimization Results}
The objective of shape optimization is to maximize the lift-to-drag ratio over a feasible region of parameters, which are $m$ and $p$ for this case. Equation \ref{eq:geo_obj} gives this objective in the form of a minimization problem, as is standard for most optimizers that perform gradient-based or gradient-free optimization. Therefore, the definition of a constrained optimization problem for airfoil is expressed as 
\begin{mini}
{m,p}{-f(m, p)}
{}{}
 \label{eq:geo_obj}
\addConstraint{m_{\min} \le m \le m_{\max}}{}
\addConstraint{p_{\min} \le p \le p_{\max}}{},
\end{mini}

\noindent where $f(m,p)$ represents ratio of lift to drag and $[m_{min}, m_{max}]$ and $[p_{min}, m_{max}]$ are bounds for feasible search region.

One of the present study's goals is to optimize the shape for any arbitrary geometry. Therefore, we integrated the DeepONet-based surrogate model with Dakota, which is a multilevel parallel object-oriented framework for
design optimization, parameter estimation, uncertainty
quantification, and sensitivity analysis \cite{osti_1868142}. Dakota is freely available and offers a very efficient and scalable implementation. We integrated the DeepONet with Dakota in a modular approach as shown in Algorithm \ref{alg: DDI}, where $\mathcal{D}$ is an algorithm chosen from a set of optimizers provided by Dakota and DeepONet-based model $\Phi$ is passed as an argument to $\mathcal{D}$. For example, to achieve the solution of Equation \eqref{eq:geo_obj}, we use an efficient global algorithm (EGO), which is a derivative-free approach that uses a Gaussian process model for the optimization of the expected improvement function and is based on the NCSU Direct algorithm \cite{finkel2004convergence}. The reason behind choosing this method is to avoid tuning various hyperparameters. To use the algorithm to solve the problem in Equation \eqref{eq:geo_obj}, we set a seed, which is to be used for Latin Hypercube Sampling (LHS) to generate the initial set of points for constructing the initial Gaussian process. To gain efficiency, we used  batch-sequential parallelization offered by Dakota on an eight-core CPU (2.3 GHz Intel core i9).

\begin{algorithm}
\caption{Integration of DeepONet-based surrogate model with Dakota}
\label{alg: DDI}
    \begin{algorithmic}
        \Require $m$, $p$, $x$, $y$:  maximum camber, the position of maximum camber, spatial coordinates for flow-field prediction
        \Require Trained {\rm DeepONet}: $\mathcal{G}(m, p, x, y)$
        \Require $\mathcal{D}$ : $\mathcal{D} \in {\text{Algorithms in Dakota}}$
        \Require $\Psi(u, v, \rho, p, \xi_g)$: Function producing the lift $L$ and drag $D$
        \Require $N$: Number of objective function evaluations
        \Require $m_{\min}, m_{\max}, p_{min}, p_{\max}$: Bounds of feasible region
        \State $n \gets 0$, $m_{\text{opt}} \gets m_0$, $p_{\text{opt}} \gets p_0$
        \Comment{Initialize}
        \While{$n~!=~N$}
           \State $m_{\text{opt}}, p_{\text{opt}} \gets \mathcal{D}(-f(m_{\text{opt}}, p_{\text{opt}}),m_{\min}, m_{\max}, p_{min}, p_{\max},\mathcal{G}, \Psi )$
           \Comment{Optimization process for parameters} 
           \State $n \gets n + 1$
        \EndWhile
    \end{algorithmic}
\end{algorithm}

\begin{figure}
\centering
\includegraphics[scale=0.5]{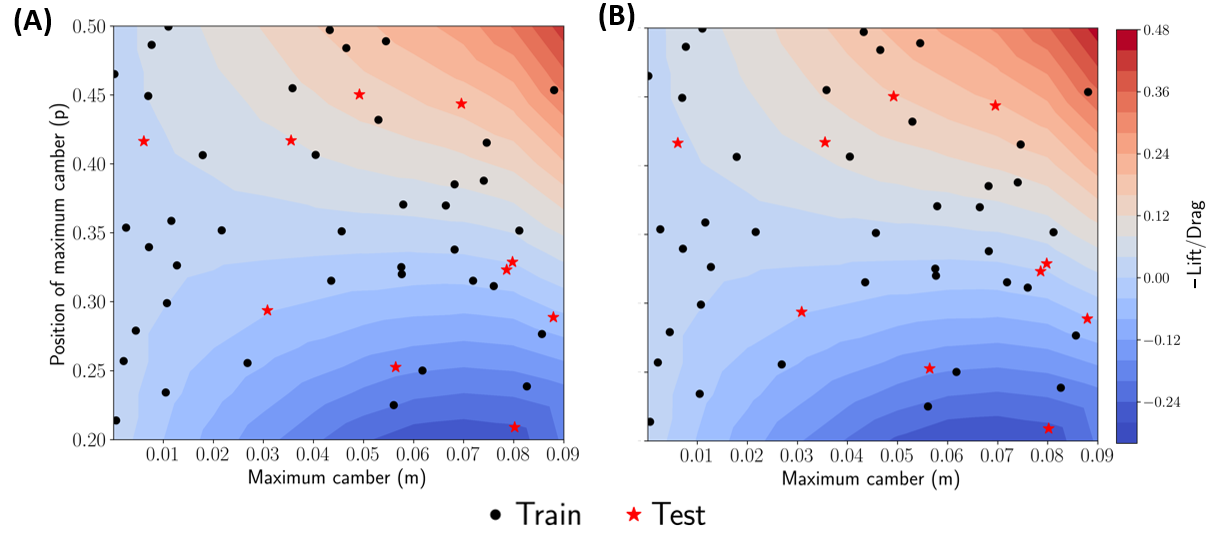}
  \caption{Visualization of the lift/drag landscape \textmd{obtained from brute force sampling of $p$ and $m$ using a $10 \times 10$ grid.} \textmd{The train and test sets are also plotted to show the sparse dataset used by the DeepONet} (A) \textmd{Landscape obtained using finite difference approximation.} (B) \textmd{Landscape obtained using automatic differentiation approximation.}}
  \label{fig:optim_landscape}
\end{figure}

The optimization landscapes for approaches (A) and (B) are shown in Fig. \ref{fig:optim_landscape} obtained by brute force evaluation of the respective objectives. Also plotted are the locations of the dataset in the parameter space $(p,m) \in [0.2, 0.5] \times [0.0, 0.09]$, displaying the sparse sampling used to obtain accurate optimization results. We also observe that the landscape for this experimental setup is simple and convex. In future work, more complex conditions like local morphing will make the landscape more complex, likely nonconvex, requiring  sophisticated optimization methods. While not needed here, we still utilize state-of-the-art optimization in our framework with Dakota. Additionally, we can see that the local minimum given the constrained parameter bounds, set to the dataset sampling bounds, is at the edge. This implies that the global minimum lies outside of our trained bounds, which is not known a priori when generating data. In future work, we hope to evaluate DeepONets efficacy in extrapolating outside the trained bounds, potentially using transfer learning.

The optimization process finds the minimizer of function in $15$ evaluations, and the optimum value which maximizes $L/D$ is $(m_*, p_*)=(0.2, 0.067)$. To achieve consistency in the optimization process, we ran the EGO algorithm 15 times with different seeds and we observed the same optimal point. Furthermore, we validated the optimization results with a comparison to other approaches in Appendix \ref{sec:appendix_optimization}. The results obtained from all the approaches are in excellent agreement and reported in detail in Table \ref{tb:geo_optim} along with their wall clock times. 
Finally, the most significant contribution of the paper is shown in Table \ref{tb:cost_compar}. As we can see, the integration of DeepONet into an airfoil geometry optimization framework has lowered the online cost of new objective evaluations by $32,000+$ times. This makes it entirely possible to have almost real-time optimization results, costing a few minutes instead of days. Furthermore, the trained models can be put on any hardware, such as a standard laptop, and real-time accurate flowfields can be predicted in seconds, meaning the geometry optimization is not hardware dependent at test time. We have demonstrated that integrating DeepONets into a geometry optimization pipeline suffers little in accuracy and provides the tradeoff of obtaining and training on an offline dataset with almost instantaneous optimization results when used online compared to a traditional CFD method.

\begin{table} 
	\centering	
	\begin{tabular}[c]{l | c}
	\toprule
        Model Type & Relative Cost of Single Objective Function Evaluation\\
        \hline
        Baseline CFD (Nektar++ placeholder) & $32,253$ \\
        DeepONet (\textbf{A}) & $1.34$ \\
        DeepONet (\textbf{B}) & $1$ \\
	\bottomrule
	\end{tabular}
    \caption{\textmd{Relative cost of single objective function evaluation during geometry optimization} A \textmd{Flowfield mapping with finite-difference approximation.} B \textmd{Flowfield mapping with automatic-differentiation approximation.}} \label{tb:cost_compar}

\end{table}

\begin{figure} 
\centering
\includegraphics[scale=0.5]{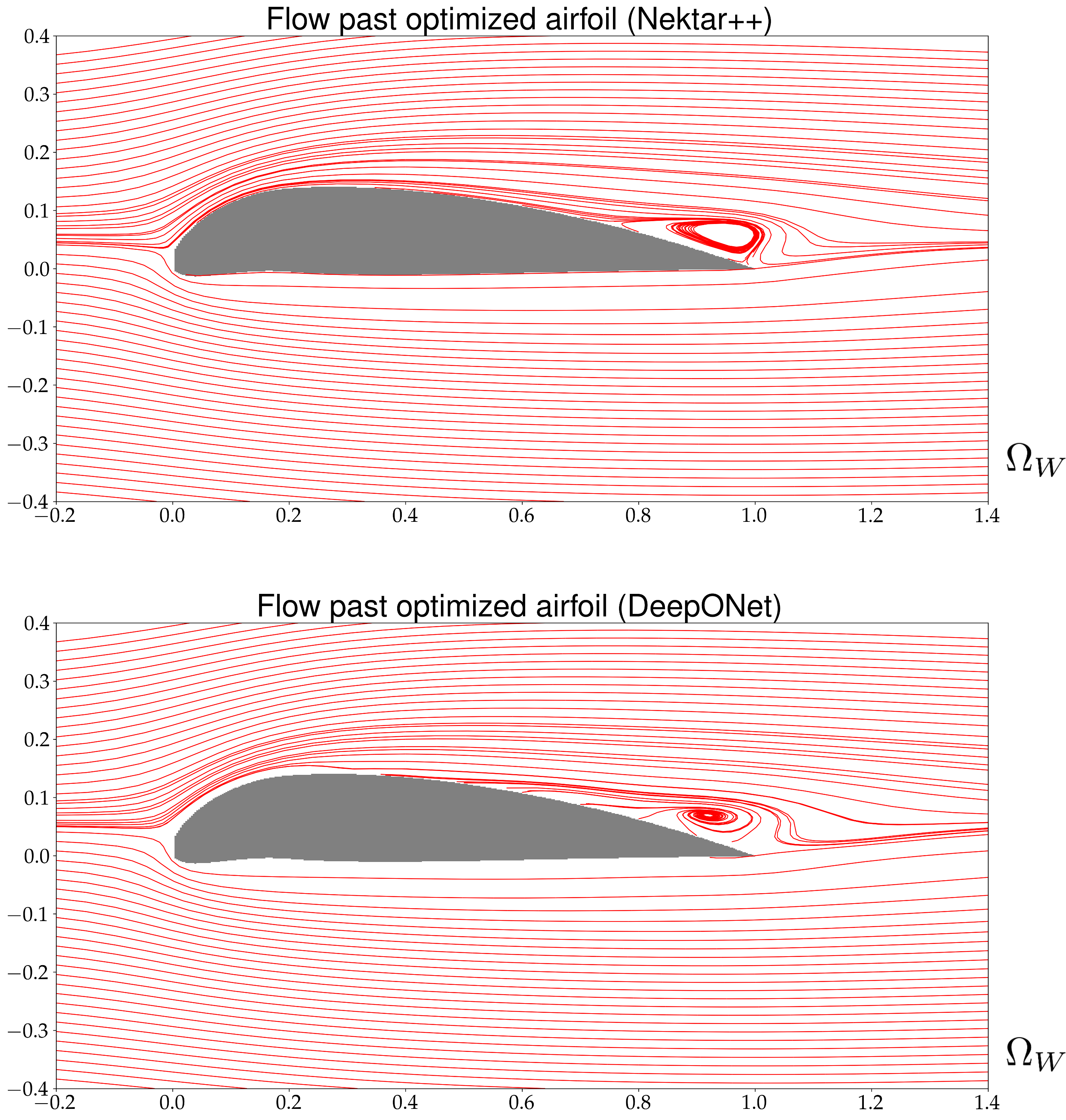}
  \caption{\textbf{Optimized airfoil. } \textmd{The streamline plot represents the flow past the optimized airfoil with $(p,m)=(0.2, 0.067)$ in $\Omega_W$. The velocity fields simulated in Nektar++ and predicted by the DeepONets are shown here.}}
  \label{fig: optimized_airfoil}
\end{figure}

 To validate the parameters of optimized airfoil $(p=0.2, 0.067)$, we compare the streamline plots in Fig. \ref {fig: optimized_airfoil} constructed using the flowfields $(u, v)$ obtained from trained DeepONet and Nektar++. In general, the streamline plots show an excellent agreement and therefore validate the workflow of shape optimization presented in this work. DeepONet is able to successfully detect and predict the circulation near the trailing edge of the airfoil. However, a closer look reveals non-physical streamlines originating from the surface of the airfoil. This is due to the error in the velocity fields near the airfoil surface, predicted by the DeepONet. The results can be further improved by giving more importance to the region near the airfoil surface (via proper weighting) during training of the DeepONet.  


%% file: conclusion.tex
\section{Conclusions}
We have successfully integrated DeepONets as a surrogate model into the shape optimization framework for airfoils. Having summarized prior work in this field, we empirically demonstrate the efficacy of DeepONets in terms of retaining sufficient flowfield accuracy used in evaluating the objective function of lift-to-drag, as well as the significant computational speed up as a replacement for a traditional CFD solver during online geometry optimization. We have provided thorough validation of the results presented as well as extensive experimentation such as two approaches (A) and (B) when approximating the wall shear stress and two forms of DeepONet inputs (NURBS and $\xi_g$) to ensure a robust pipeline. Importantly,  DeepONets exhibit almost no generalization error over the dataset, so it follows that the resulting optimized geometry ($p = 0.2, m = 0.067$) is accurate and achieved $32,253$ speed-up compared to the CFD baseline. The framework is general and can address more complex problems with multiple inputs, e.g. different Mach numbers and different angles of attack that can be inputs to either the branch or the trunk networks. Hence, with relatively small  modifications, such a framework can handle optimization in the high speed flow regimes that exhibit flow unsteadiness, shocks, non-equilibrium chemistry, and even morphing geometry. Furthermore, the approaches presented are flexible due to the integration of machine learning in the form of function-to-function maps using DeepONet. Therefore, improvements such as the introduction of multi-fidelity training and physics-informed machine learning can be leveraged to reduce the cost of data generation. We also successfully show the application of automatic differentiation, which performs comparably to the traditional approach of finite differences in the wall shear stress calculation. Finally, we hope to utilize transfer learning and uncertainty quantification using the recently developed library {\em NeuralUQ} \cite{zou2022neuraluq} to extrapolate outside of the trained geometric parameter domain to find global optima with confidence.

%% file: appendix.tex
\section*{Appendix}

\subsection{Nektar++ cross-verification}
\begin{figure}[H]
  \begin{center}
\includegraphics[scale=0.4]{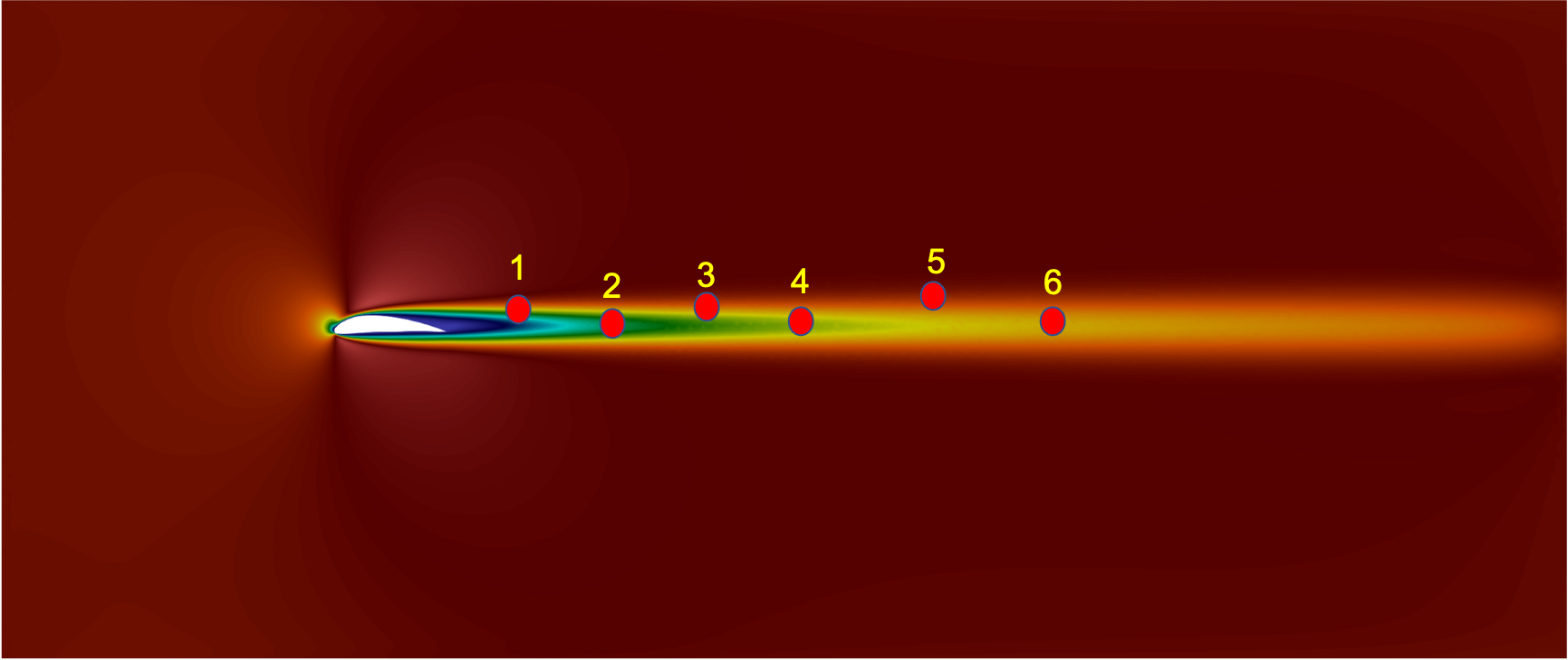}
\caption{\textmd{History points locations in the wake for NACA airfoils. The conservative variable values are stored in time at these locations to monitor the steady-state solution.}}

    \label{fig:historypoints}
  \end{center}
  
\end{figure}

\begin{figure}  
  \begin{center}
    \begin{tabular}{cc}
    \includegraphics[scale=0.2]{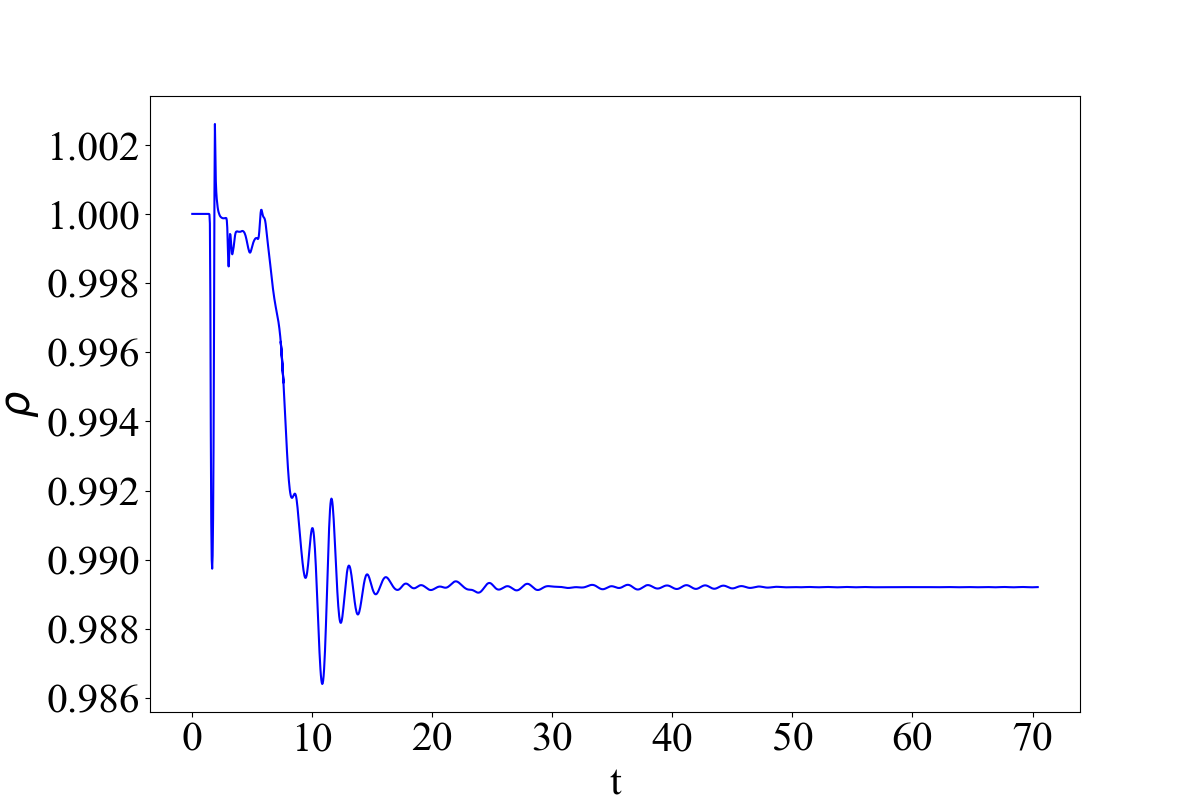}
 &
    \includegraphics[scale=0.2]{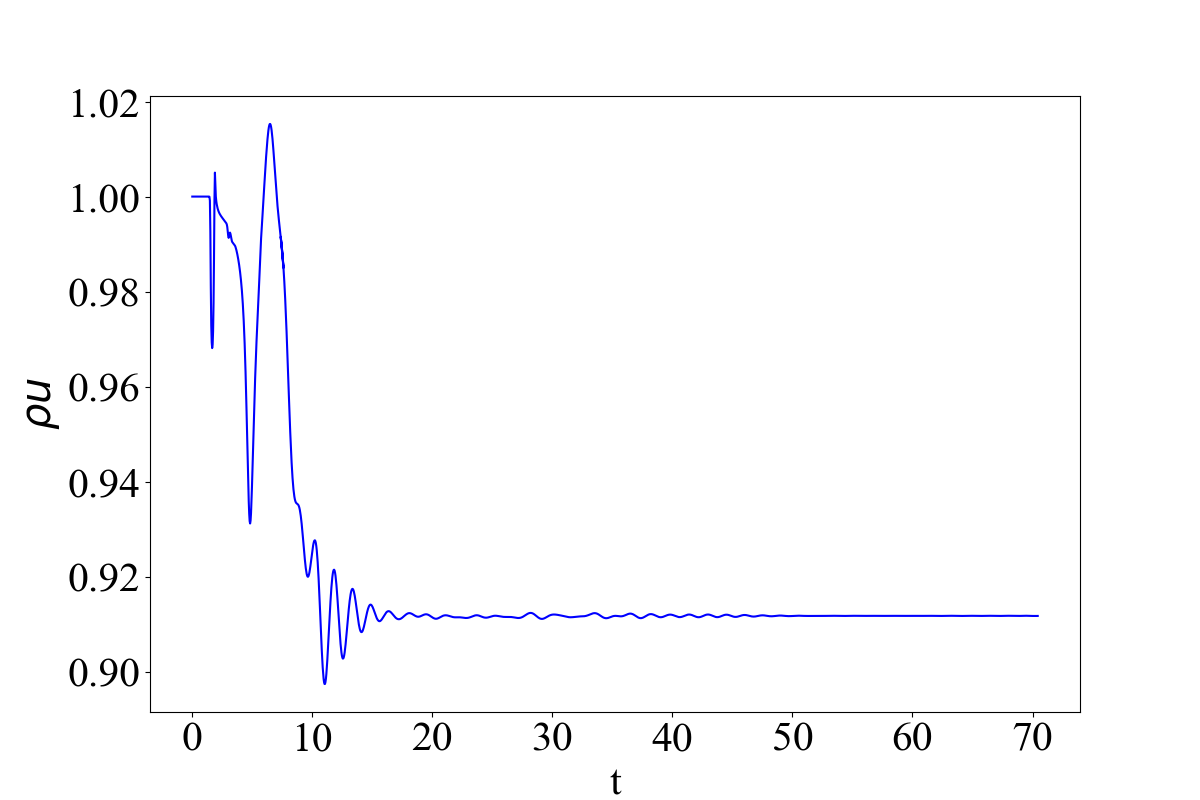}  
  \\
  (a) Density & (b) x-momentum
  \\
     \includegraphics[scale=0.2]{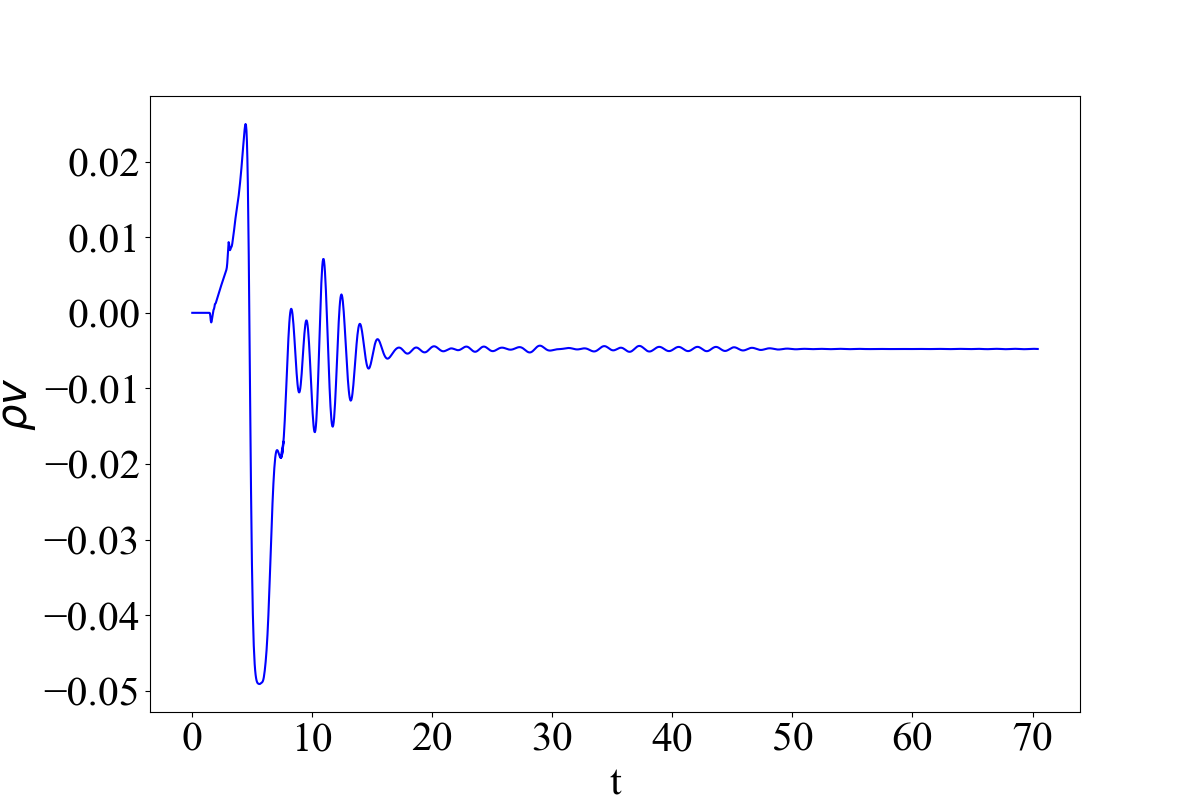}
 &
    \includegraphics[scale=0.2]{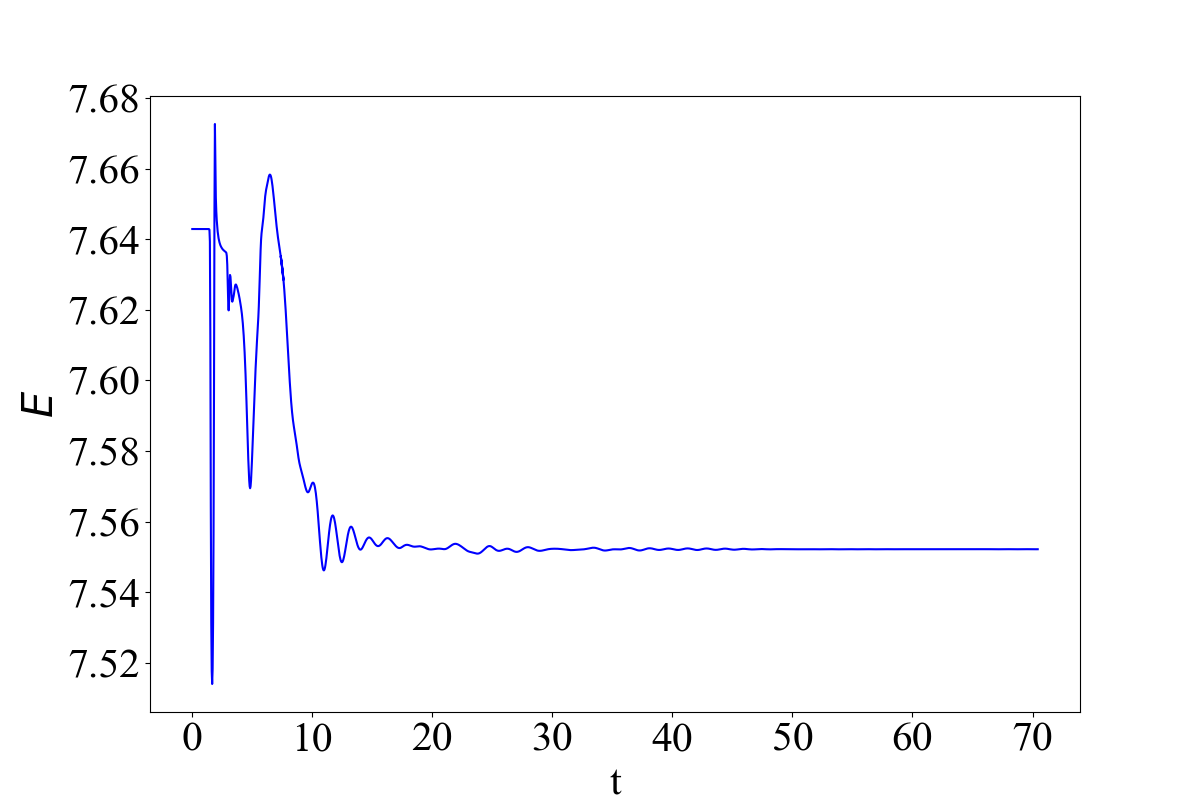}
  \\
  (c) y-momentum & (d) Total energy
\end{tabular} 

\caption{\textmd{Time history of (a) $\rho$, (b) $\rho u$, (c) $\rho v$, and (d) $E$ at point 5 (see Fig.~\ref{fig:historypoints}) for the NACA7315 airfoil. The profiles show that the simulation has reached a steady-state solution where the flow variables reach constant states.}}

    \label{fig:timehist}
  \end{center}
  
\end{figure}

To ensure a steady-state solution, we selected the flow parameters as $M=0.5$ and  $Re=500$. Calculating the steady-state solution requires careful consideration. To ensure the steady state solution, for each NACA profile, we recorded the value of conservative variables at six different locations in the wake. We then examined the conservative variables' time history to ensure the steady state is reached and the solution update has stopped. Fig.~\ref{fig:historypoints} shows the location of the history points, where the transient flow could last longer than other spatial locations. As a sample, we plot the time history of variables at point 5 (Fig.~\ref{fig:timehist}), which experiences the highest flow fluctuations in time. According to Fig.~\ref{fig:timehist}, the solution reached a stationary state when the conservative variables approached constant values. 

\begin{figure}  
  \begin{center}
    \begin{tabular}{cc}
     \includegraphics[scale=0.2,trim={2cm 8cm 0 5cm},clip]{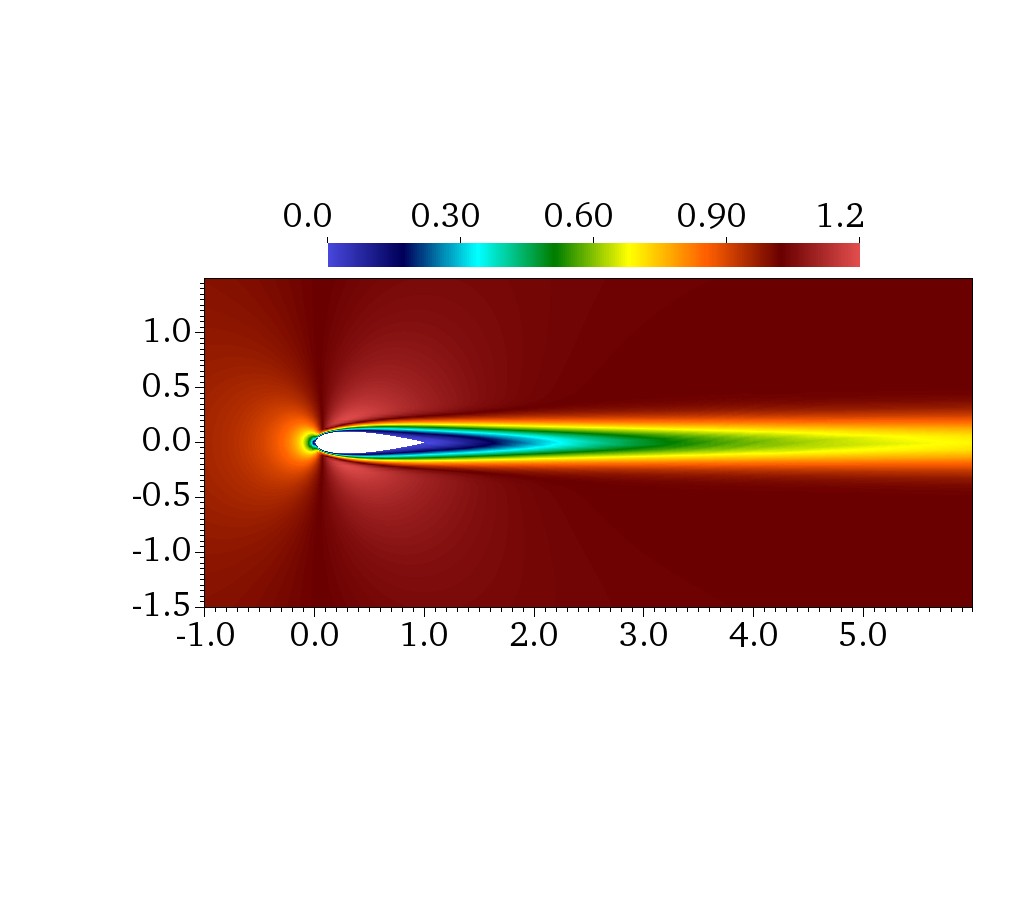}
 &
    \includegraphics[scale=0.2,trim={2cm 8cm 0 5cm},clip]{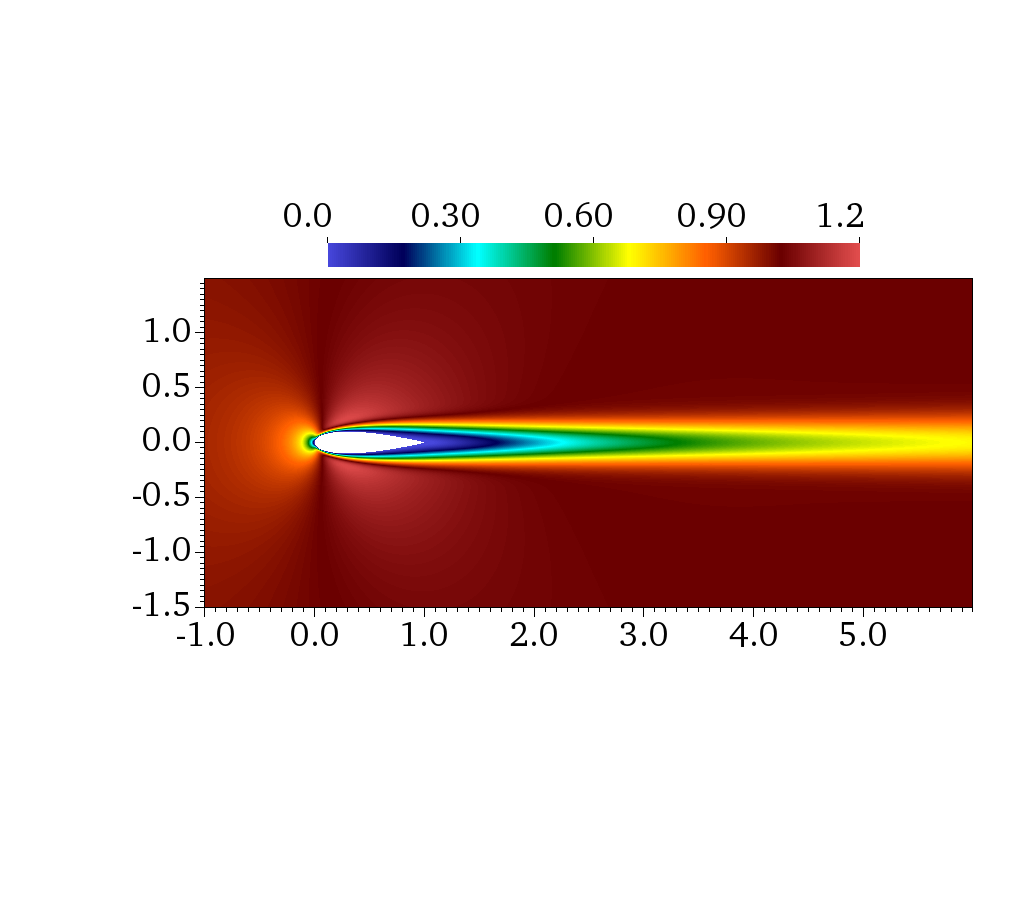}
    
  \\
  (a) x-Velocity Nektar++& (b) x-Velocity DSEM
  \\
     \includegraphics[scale=0.2,trim={2cm 8cm 0 5cm},clip]{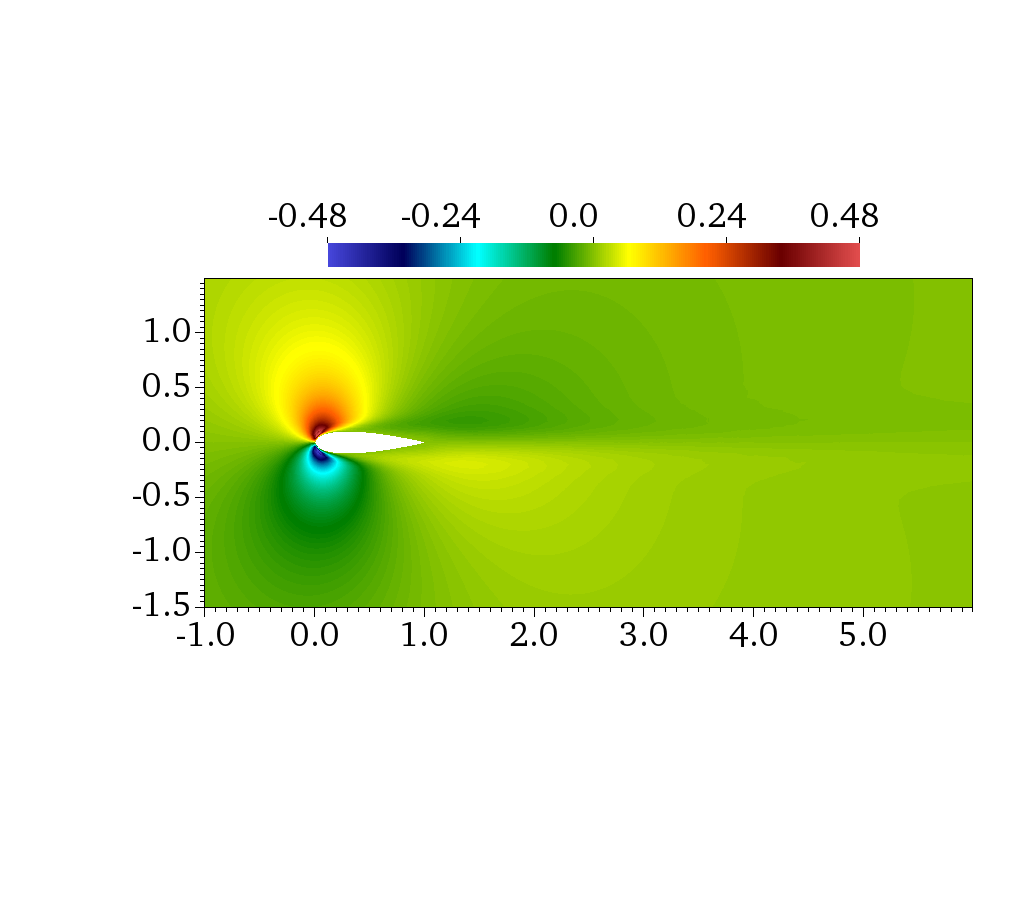}
 &
    \includegraphics[scale=0.2,trim={2cm 8cm 0 5cm},clip]{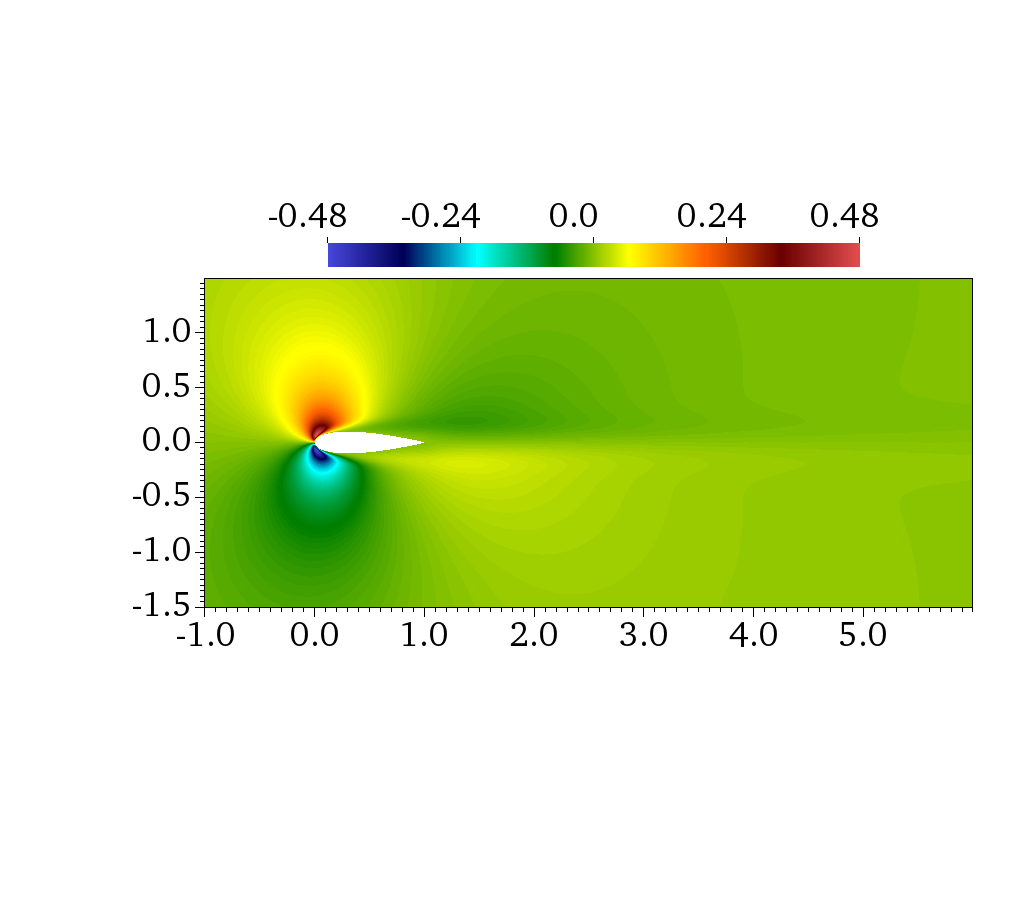}
  \\
  (c) y-Velocity Nektar++& (d) y-Velocity DSEM
    \\
     \includegraphics[scale=0.2,trim={2cm 8cm 0 5cm},clip]{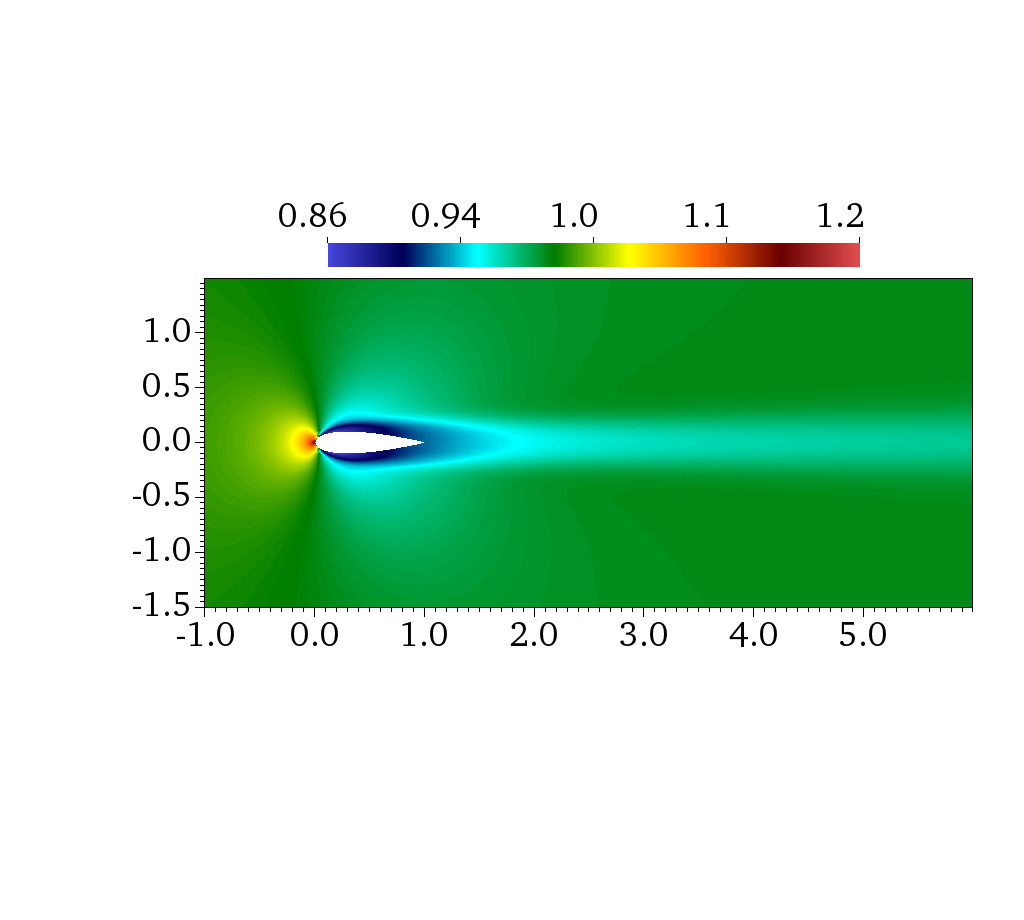}
 &
    \includegraphics[scale=0.2,trim={2cm 8cm 0 5cm},clip]{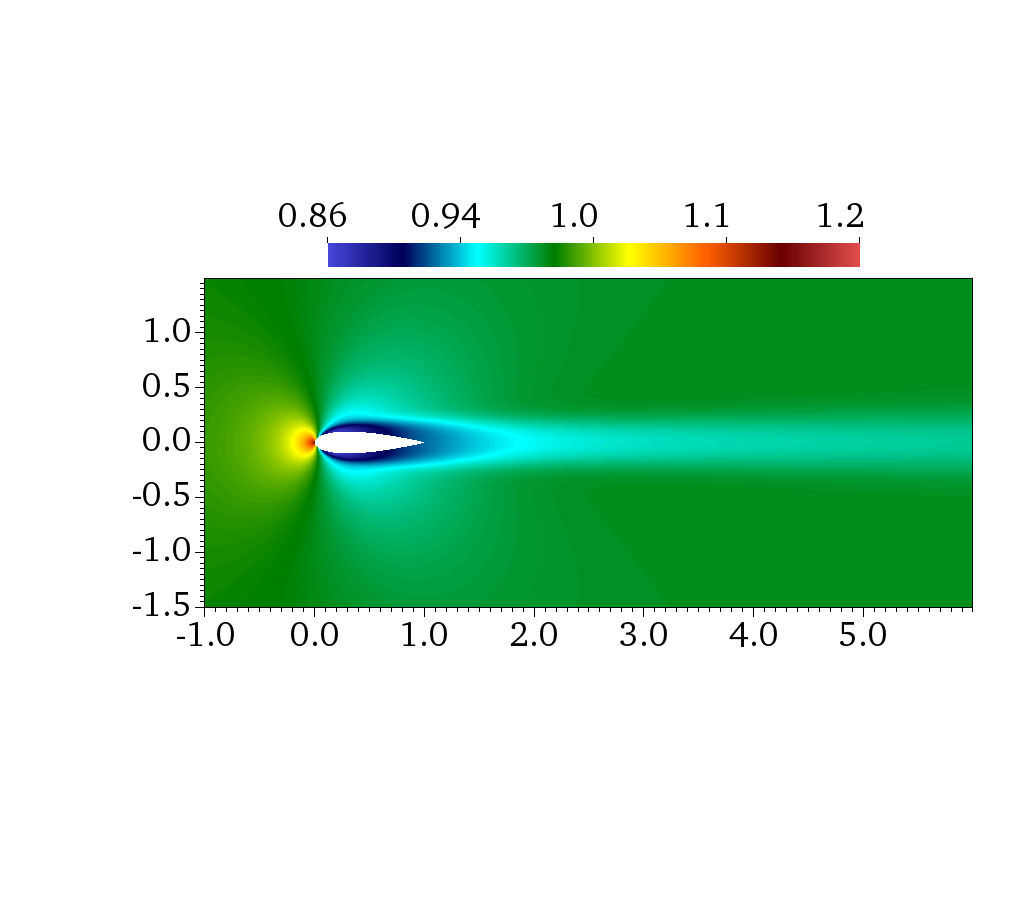}
  \\
  (e) Density Nektar++& (f) Density DSEM
      \\
     \includegraphics[scale=0.2,trim={2cm 8cm 0 5cm},clip]{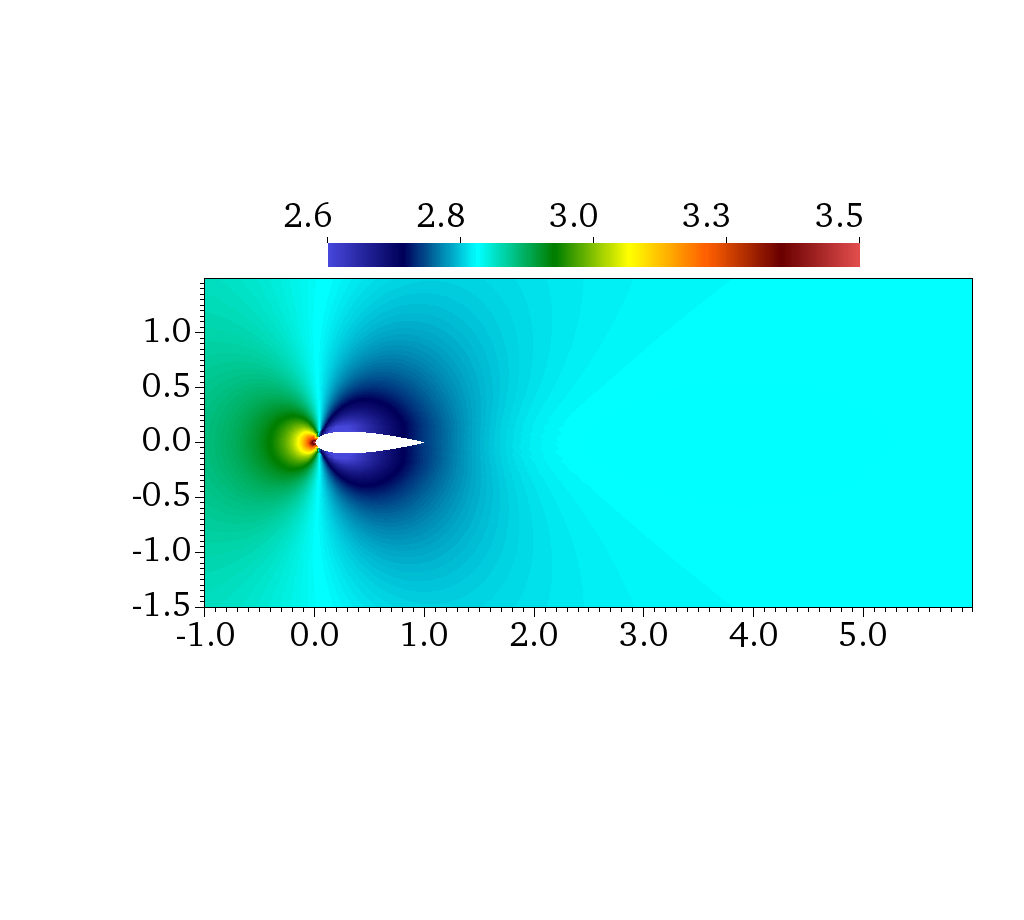}
 &
    \includegraphics[scale=0.2,trim={2cm 8cm 0 5cm},clip]{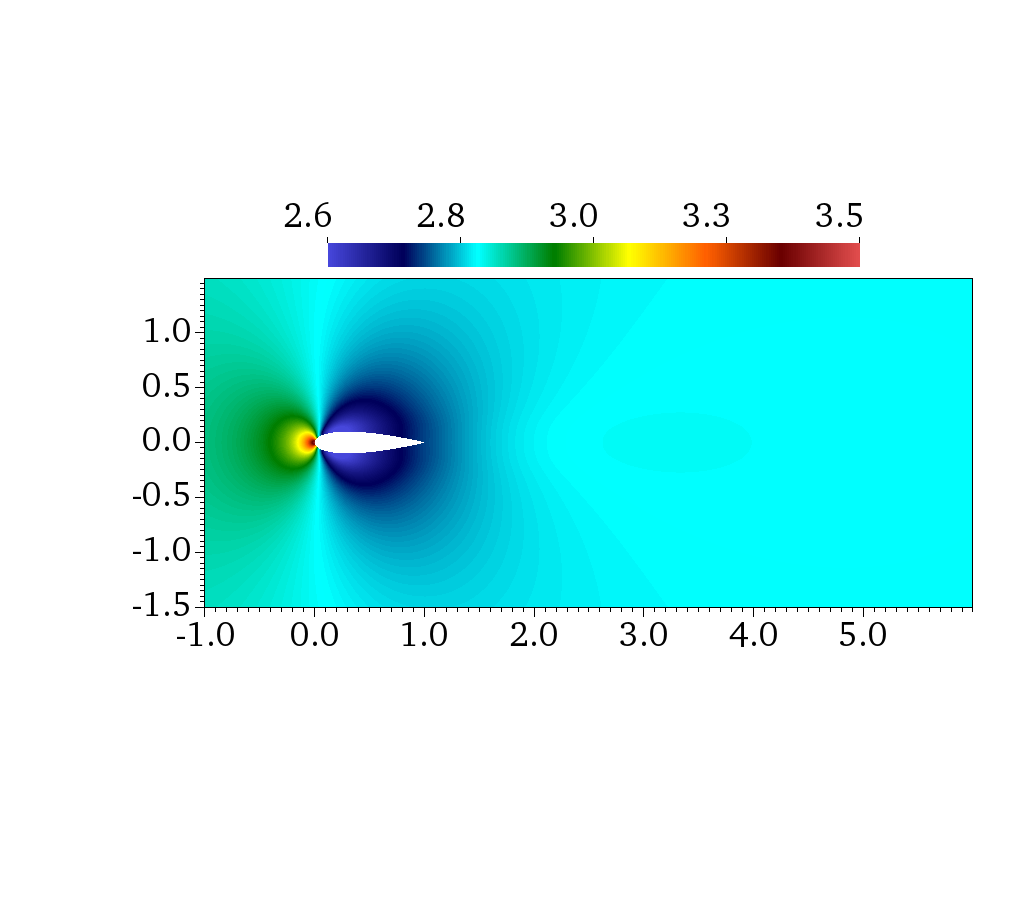}
  \\
  (g) Pressure Nektar++& (h) Pressure DSEM
\end{tabular} 

\caption{\textmd{Cross verification of steady state flow around NACA0020 obtained by the DSEM code and Nektar++. The flow parameters are set as $Re=500$ and $M=0.5$.}}

    \label{fig:crossveri}
  \end{center}
  
\end{figure}

We also validated the simulation setup of NACA airfoils in NekTar++ with the results obtained by an in-house code based on the discontinuous spectral element method of Kopriva \cite{KOPRIVA1996244,PEYVAN2021110261}. We selected the NACA0020 airfoil for cross-verification. The steady-state solutions of the flow with $M=0.5$ and $Re=500$ are computed, and the results are shown in Fig.~\ref{fig:crossveri}. The flow field primitive variables computed by both solvers agree and show the validity of the NekTar++ simulation setup, including mesh and simulation parameters. According to Fig.~\ref{fig:crossveri}, the number of elements employed for the NekTar++ simulations is sufficient for an accurate prediction. After validating the flow field, we performed an extra flow simulation around the NACA4402 airfoil. The drag and lift coefficients of this airfoil are reported by Kunz \cite{kunz2003aerodynamics} but for an incompressible flow at $Re=1000$. 

We employed the automatic mesh generation setup used for the training set to create the mesh and used the same simulation setup as the training set. We computed the drag and lift coefficients and compared them with the literature. Table.~\ref{tb:liftdrag} compares the drag and lift coefficients computed by Nektar++ with values reported by Kunz \cite{kunz2003aerodynamics} for a similar but not exactly the same set up.

\begin{table}[H]  
\centering
\begin{tabular}{lccc}
\hline
    &$C_d$&$C_l$& Error\\
    \hline
    Nektar++&0.1050&0.1700&$1.7
\%$\\
    Kunz \cite{kunz2003aerodynamics} &0.1032&0.1852&$8.2\%$\\
    \hline
\end{tabular}
\caption{\textmd{Comparison between the drag and lift coefficients computed by Nektar++ with Kunz \cite{kunz2003aerodynamics}. For the reference solution, the setup is different, i.e., the Reynolds number is $1000$, and the flow is assumed to be incompressible.}} \label{tb:liftdrag}
	
\end{table}

\subsection{Geometry optimization validation}
\label{sec:appendix_optimization}

To validate the main geometry optimization findings in the manuscript using Dakota, we also evaluate the objective function using brute force and SciPy's \cite{2020SciPy-NMeth} dual-annealing method. Brute force is evaluated using a $10 \times 10$ grid on the geometric parameter space $(p,m) \in [0.2, 0.5] \times [0.0, 0.09]$, and therefore requires 100 evaluations of the objective. The dual-annealing method is set to have a maximum amount of 50 evaluations but likely could be set to fewer. As seen in Table \ref{tb:geo_optim}, all methods discover the same optimal set of parameters and $\frac{L}{D}$, regardless of optimizer or partial approximation given by (A) and (B). 
\begin{table}  
\centering
\begin{tabular}[c]{c|c|c|c}
	\toprule
        Model Type & Optimized Parameters (p, m) & Optimized $\frac{L}{D}$ & Computation  
        Cost (sec)\\
        \hline
        Brute Force (\textbf{A}) & (0.2, 0.070) & 0.272 & 332.63 \\
        Brute Force (\textbf{B}) & (0.2, 0.070) & 0.281 & 264.12 \\
        SciPy dual-annealing (\textbf{A}) & (0.2, 0.064) & 0.269 & 196.09 \\
        SciPy dual-annealing (\textbf{B}) & (0.2, 0.064) & 0.281 & 165.57\\
        Dakota (\textbf{A}) & (0.2, 0.067) & 0.269 & 157.71 \\
        Dakota (\textbf{B}) & (0.2, 0.063) & 0.282  & 105.00 \\
	\bottomrule
	\end{tabular}
    \caption{\textmd{Geometry optimization results for different DeepONet methodologies.} (A) \textmd{Flowfield mapping with finite-difference  approximation.} (B) \textmd{Flowfield mapping with automatic-differentiation approximation. The cost of these methods were obtained on a eight-core CPU (2.3 GHz Intel core i9) with 16 GB 2667 MHz DDR4 of MacBook Pro 2019.}} \label{tb:geo_optim}
\end{table}

%% file: main.bbl
\begin{thebibliography}{10}

\bibitem{lu2021learning}
Lu~Lu, Pengzhan Jin, Guofei Pang, Zhongqiang Zhang, and George~Em Karniadakis.
\newblock Learning nonlinear operators via deeponet based on the universal
  approximation theorem of operators.
\newblock {\em Nature Machine Intelligence}, 3(3):218--229, 2021.

\bibitem{lu2019deeponet}
Lu~Lu, Pengzhan Jin, and George~Em Karniadakis.
\newblock Deeponet: Learning nonlinear operators for identifying differential
  equations based on the universal approximation theorem of operators.
\newblock {\em arXiv preprint arXiv:1910.03193}, 2019.

\bibitem{raissi2019physics}
Maziar Raissi, Paris Perdikaris, and George~E Karniadakis.
\newblock Physics-informed neural networks: A deep learning framework for
  solving forward and inverse problems involving nonlinear partial differential
  equations.
\newblock {\em Journal of Computational physics}, 378:686--707, 2019.

\bibitem{de2022bi}
Subhayan De, Malik Hassanaly, Matthew Reynolds, Ryan~N. King, and Alireza
  Doostan.
\newblock Bi-fidelity modeling of uncertain and partially unknown systems using
  deeponets. {P}reprint at
  \hyperlink{https://arxiv.org/abs/2204.00997}{https://arxiv.org/abs/2204.00997},
  2022.

\bibitem{howard2022multifidelity}
Amanda~A. Howard, Mauro Perego, George~E. Karniadakis, and Panos Stinis.
\newblock Multifidelity deep operator networks. {P}reprint at
  \hyperlink{https://arxiv.org/abs/2204.09157}{https://arxiv.org/abs/2204.09157},
  2022.

\bibitem{lu2022multifidelity}
Lu~Lu, Rapha\"el Pestourie, Steven~G. Johnson, and Giuseppe Romano.
\newblock Multifidelity deep neural operators for efficient learning of partial
  differential equations with application to fast inverse design of nanoscale
  heat transport.
\newblock {\em Phys. Rev. Research}, 4:023210, 2022.

\bibitem{jin2022mionet}
Pengzhan Jin, Shuai Meng, and Lu~Lu.
\newblock Mionet: Learning multiple-input operators via tensor product.
\newblock {\em arXiv preprint arXiv:2202.06137}, 2022.

\bibitem{zhu2022reliable}
Min Zhu, Handi Zhang, Anran Jiao, George~Em Karniadakis, and Lu~Lu.
\newblock Reliable extrapolation of deep neural operators informed by physics
  or sparse observations.
\newblock {\em arXiv preprint arXiv:2212.06347}, 2022.

\bibitem{hesthaven2018non}
Jan~S Hesthaven and Stefano Ubbiali.
\newblock Non-intrusive reduced order modeling of nonlinear problems using
  neural networks.
\newblock {\em Journal of Computational Physics}, 363:55--78, 2018.

\bibitem{hesthaven2016certified}
Jan~S Hesthaven, Gianluigi Rozza, Benjamin Stamm, et~al.
\newblock {\em Certified reduced basis methods for parametrized partial
  differential equations}, volume 590.
\newblock Springer, 2016.

\bibitem{benner2017model}
Peter Benner, Mario Ohlberger, Anthony Patera, Gianluigi Rozza, and Karsten
  Urban.
\newblock {\em Model reduction of parametrized systems}.
\newblock Springer, 2017.

\bibitem{williams2015data}
Matthew~O Williams, Ioannis~G Kevrekidis, and Clarence~W Rowley.
\newblock A data--driven approximation of the koopman operator: Extending
  dynamic mode decomposition.
\newblock {\em Journal of Nonlinear Science}, 25(6):1307--1346, 2015.

\bibitem{chiavazzo2014reduced}
Eliodoro Chiavazzo, Charles~W Gear, Carmeline~J Dsilva, Neta Rabin, and
  Ioannis~G Kevrekidis.
\newblock Reduced models in chemical kinetics via nonlinear data-mining.
\newblock {\em Processes}, 2(1):112--140, 2014.

\bibitem{lieberman2010parameter}
Chad Lieberman, Karen Willcox, and Omar Ghattas.
\newblock Parameter and state model reduction for large-scale statistical
  inverse problems.
\newblock {\em SIAM Journal on Scientific Computing}, 32(5):2523--2542, 2010.

\bibitem{bui2008model}
Tan Bui-Thanh, Karen Willcox, and Omar Ghattas.
\newblock Model reduction for large-scale systems with high-dimensional
  parametric input space.
\newblock {\em SIAM Journal on Scientific Computing}, 30(6):3270--3288, 2008.

\bibitem{benner2015survey}
Peter Benner, Serkan Gugercin, and Karen Willcox.
\newblock A survey of projection-based model reduction methods for parametric
  dynamical systems.
\newblock {\em SIAM review}, 57(4):483--531, 2015.

\bibitem{amsallem2015design}
David Amsallem, Matthew Zahr, Youngsoo Choi, and Charbel Farhat.
\newblock Design optimization using hyper-reduced-order models.
\newblock {\em Structural and Multidisciplinary Optimization}, 51(4):919--940,
  2015.

\bibitem{carlberg2008compact}
Kevin Carlberg and Charbel Farhat.
\newblock A compact proper orthogonal decomposition basis for
  optimization-oriented reduced-order models.
\newblock In {\em 12th AIAA/ISSMO multidisciplinary analysis and optimization
  conference}, page 5964, 2008.

\bibitem{choi2020gradient}
Youngsoo Choi, Gabriele Boncoraglio, Spenser Anderson, David Amsallem, and
  Charbel Farhat.
\newblock Gradient-based constrained optimization using a database of linear
  reduced-order models.
\newblock {\em Journal of Computational Physics}, 423:109787, 2020.

\bibitem{kontolati2022influence}
Katiana Kontolati, Somdatta Goswami, Michael~D Shields, and George~Em
  Karniadakis.
\newblock On the influence of over-parameterization in manifold based
  surrogates and deep neural operators.
\newblock {\em arXiv preprint arXiv:2203.05071}, 2022.

\bibitem{yu2018influence}
Yin Yu, Zhoujie Lyu, Zelu Xu, and Joaquim~RRA Martins.
\newblock On the influence of optimization algorithm and initial design on wing
  aerodynamic shape optimization.
\newblock {\em Aerospace Science and Technology}, 75:183--199, 2018.

\bibitem{reuther1996aerodynamic}
James Reuther, Antony Jameson, James Farmer, Luigi Martinelli, and David
  Saunders.
\newblock Aerodynamic shape optimization of complex aircraft configurations via
  an adjoint formulation.
\newblock In {\em 34th aerospace sciences meeting and exhibit}, page~94, 1996.

\bibitem{carpentieri2007adjoint}
Giampietro Carpentieri, Barry Koren, and Michel~JL van Tooren.
\newblock Adjoint-based aerodynamic shape optimization on unstructured meshes.
\newblock {\em Journal of Computational Physics}, 224(1):267--287, 2007.

\bibitem{nadarajah2001studies}
Siva Nadarajah and Antony Jameson.
\newblock Studies of the continuous and discrete adjoint approaches to viscous
  automatic aerodynamic shape optimization.
\newblock In {\em 15th AIAA computational fluid dynamics conference}, page
  2530, 2001.

\bibitem{srinath2010adjoint}
DN~Srinath and Sanjay Mittal.
\newblock An adjoint method for shape optimization in unsteady viscous flows.
\newblock {\em Journal of Computational Physics}, 229(6):1994--2008, 2010.

\bibitem{chernukhin2013multimodality}
Oleg Chernukhin and David~W Zingg.
\newblock Multimodality and global optimization in aerodynamic design.
\newblock {\em AIAA journal}, 51(6):1342--1354, 2013.

\bibitem{li2019surrogate}
Jichao Li, Jinsheng Cai, and Kun Qu.
\newblock Surrogate-based aerodynamic shape optimization with the active
  subspace method.
\newblock {\em Structural and Multidisciplinary Optimization}, 59(2):403--419,
  2019.

\bibitem{wu2022aerodynamic}
Xiaojing Wu, Zijun Zuo, and Long Ma.
\newblock Aerodynamic data-driven surrogate-assisted teaching-learning-based
  optimization (tlbo) framework for constrained transonic airfoil and wing
  shape designs.
\newblock {\em Aerospace}, 9(10):610, 2022.

\bibitem{eberhart1995new}
Russell Eberhart and James Kennedy.
\newblock A new optimizer using particle swarm theory.
\newblock In {\em MHS'95. Proceedings of the sixth international symposium on
  micro machine and human science}, pages 39--43. Ieee, 1995.

\bibitem{krige1951statistical}
Daniel~G Krige.
\newblock A statistical approach to some basic mine valuation problems on the
  witwatersrand.
\newblock {\em Journal of the Southern African Institute of Mining and
  Metallurgy}, 52(6):119--139, 1951.

\bibitem{liu2017efficient}
J~Liu, W-P Song, Z-H Han, and Y~Zhang.
\newblock Efficient aerodynamic shape optimization of transonic wings using a
  parallel infilling strategy and surrogate models.
\newblock {\em Structural and Multidisciplinary Optimization}, 55(3):925--943,
  2017.

\bibitem{lepine2001optimized}
Jerome Lepine, Francois Guibault, Jean-Yves Trepanier, and Francois Pepin.
\newblock Optimized nonuniform rational b-spline geometrical representation for
  aerodynamic design of wings.
\newblock {\em AIAA journal}, 39(11):2033--2041, 2001.

\bibitem{wang2019adjoint}
Kun Wang, Shengjiao Yu, Zheng Wang, Renzhong Feng, and Tiegang Liu.
\newblock Adjoint-based airfoil optimization with adaptive isogeometric
  discontinuous galerkin method.
\newblock {\em Computer Methods in Applied Mechanics and Engineering},
  344:602--625, 2019.

\bibitem{papadimitriou2016aerodynamic}
Dimitrios~I Papadimitriou and Costas Papadimitriou.
\newblock Aerodynamic shape optimization for minimum robust drag and lift
  reliability constraint.
\newblock {\em Aerospace Science and Technology}, 55:24--33, 2016.

\bibitem{hicks1978wing}
Raymond~M Hicks and Preston~A Henne.
\newblock Wing design by numerical optimization.
\newblock {\em Journal of Aircraft}, 15(7):407--412, 1978.

\bibitem{painchaud2006airfoil}
Simon Painchaud-Ouellet, Christophe Tribes, Jean-Yves Tr{\'e}panier, and
  Dominique Pelletier.
\newblock Airfoil shape optimization using a nonuniform rational b-splines
  parametrization under thickness constraint.
\newblock {\em AIAA journal}, 44(10):2170--2178, 2006.

\bibitem{chen2017airfoil}
Guodong Chen and Krzysztof Fidkowski.
\newblock Airfoil shape optimization using output-based adapted meshes.
\newblock In {\em 23rd AIAA Computational Fluid Dynamics Conference}, page
  3102, 2017.

\bibitem{he2019robust}
Xiaolong He, Jichao Li, Charles~A Mader, Anil Yildirim, and Joaquim~RRA
  Martins.
\newblock Robust aerodynamic shape optimization—from a circle to an airfoil.
\newblock {\em Aerospace Science and Technology}, 87:48--61, 2019.

\bibitem{wu2019benchmark}
Xiaojing Wu, Weiwei Zhang, Xuhao Peng, and Ziyi Wang.
\newblock Benchmark aerodynamic shape optimization with the pod-based cst
  airfoil parametric method.
\newblock {\em Aerospace Science and Technology}, 84:632--640, 2019.

\bibitem{akram2021cfd}
Md~Tausif Akram and Man-Hoe Kim.
\newblock Cfd analysis and shape optimization of airfoils using class shape
  transformation and genetic algorithm—part i.
\newblock {\em Applied Sciences}, 11(9):3791, 2021.

\bibitem{zhang2021multi}
Xinshuai Zhang, Fangfang Xie, Tingwei Ji, Zaoxu Zhu, and Yao Zheng.
\newblock Multi-fidelity deep neural network surrogate model for aerodynamic
  shape optimization.
\newblock {\em Computer Methods in Applied Mechanics and Engineering},
  373:113485, 2021.

\bibitem{zhiwei2020non}
SUN Zhiwei, WANG Chen, Yu~Zheng, BAI Junqiang, LI~Zheng, XIA Qiang, and
  FU~Qiujun.
\newblock Non-intrusive reduced-order model for predicting transonic flow with
  varying geometries.
\newblock {\em Chinese Journal of Aeronautics}, 33(2):508--519, 2020.

\bibitem{renganathan2021enhanced}
S~Ashwin Renganathan, Romit Maulik, and Jai Ahuja.
\newblock Enhanced data efficiency using deep neural networks and gaussian
  processes for aerodynamic design optimization.
\newblock {\em Aerospace Science and Technology}, 111:106522, 2021.

\bibitem{du2021rapid}
Xiaosong Du, Ping He, and Joaquim~RRA Martins.
\newblock Rapid airfoil design optimization via neural networks-based
  parameterization and surrogate modeling.
\newblock {\em Aerospace Science and Technology}, 113:106701, 2021.

\bibitem{liao2021multi}
Peng Liao, Wei Song, Peng Du, and Hang Zhao.
\newblock Multi-fidelity convolutional neural network surrogate model for
  aerodynamic optimization based on transfer learning.
\newblock {\em Physics of Fluids}, 33(12):127121, 2021.

\bibitem{tao2019application}
Jun Tao and Gang Sun.
\newblock Application of deep learning based multi-fidelity surrogate model to
  robust aerodynamic design optimization.
\newblock {\em Aerospace Science and Technology}, 92:722--737, 2019.

\bibitem{gmsh}
Christophe Geuzaine and Jean-Francois Remacle.
\newblock Gmsh.

\bibitem{CANTWELL2015205}
C.D. Cantwell, D.~Moxey, A.~Comerford, A.~Bolis, G.~Rocco, G.~Mengaldo, D.~{De
  Grazia}, S.~Yakovlev, J.-E. Lombard, D.~Ekelschot, B.~Jordi, H.~Xu,
  Y.~Mohamied, C.~Eskilsson, B.~Nelson, P.~Vos, C.~Biotto, R.M. Kirby, and S.J.
  Sherwin.
\newblock Nektar++: An open-source spectral/hp element framework.
\newblock {\em Computer Physics Communications}, 192:205--219, 2015.

\bibitem{MOXEY2020107110}
David Moxey, Chris~D. Cantwell, Yan Bao, Andrea Cassinelli, Giacomo
  Castiglioni, Sehun Chun, Emilia Juda, Ehsan Kazemi, Kilian Lackhove, Julian
  Marcon, Gianmarco Mengaldo, Douglas Serson, Michael Turner, Hui Xu, Joaquim
  Peiró, Robert~M. Kirby, and Spencer~J. Sherwin.
\newblock Nektar++: Enhancing the capability and application of high-fidelity
  spectral/hp element methods.
\newblock {\em Computer Physics Communications}, 249:107110, 2020.

\bibitem{jacobs_ward_pinkerton_1933}
Eastman~N Jacobs, Kenneth~E Ward, and Robert~M Pinkerton.
\newblock The characteristics of 78 related airfoil sections from tests in the
  variable-density wind tunnel.
\newblock {\em National Advisory Committee for Aeronautics}, 1933.

\bibitem{bingol2019geomdl}
Onur~Rauf Bingol and Adarsh Krishnamurthy.
\newblock {NURBS-Python}: An open-source object-oriented {NURBS} modeling
  framework in {Python}.
\newblock {\em {SoftwareX}}, 9:85--94, 2019.

\bibitem{mark2008computational}
de~Berg Mark, Cheong Otfried, van~Kreveld Marc, and Overmars Mark.
\newblock {\em Computational geometry algorithms and applications}.
\newblock Spinger, 2008.

\bibitem{mengaldo2014guide}
Gianmarco Mengaldo, Daniele De~Grazia, Freddie Witherden, Antony Farrington,
  Peter Vincent, Spencer Sherwin, and Joaquim Peiro.
\newblock A guide to the implementation of boundary conditions in compact
  high-order methods for compressible aerodynamics.
\newblock In {\em 7th AIAA Theoretical Fluid Mechanics Conference}, page 2923,
  2014.

\bibitem{chen1995universal}
Tianping Chen and Hong Chen.
\newblock Universal approximation to nonlinear operators by neural networks
  with arbitrary activation functions and its application to dynamical systems.
\newblock {\em IEEE Transactions on Neural Networks}, 6(4):911--917, 1995.

\bibitem{li2020neural}
Zongyi Li, Nikola Kovachki, Kamyar Azizzadenesheli, Burigede Liu, Kaushik
  Bhattacharya, Andrew Stuart, and Anima Anandkumar.
\newblock Neural operator: Graph kernel network for partial differential
  equations, 2020.

\bibitem{baydin2018automatic}
Atilim~Gunes Baydin, Barak~A Pearlmutter, Alexey~Andreyevich Radul, and
  Jeffrey~Mark Siskind.
\newblock Automatic differentiation in machine learning: a survey.
\newblock {\em Journal of Marchine Learning Research}, 18:1--43, 2018.

\bibitem{osti_1868142}
Brian Adams, William Bohnhoff, Keith Dalbey, Mohamed Ebeida, John Eddy, Michael
  Eldred, Russell Hooper, Patricia Hough, Kenneth Hu, John Jakeman, Mohammad
  Khalil, Kathryn Maupin, Jason~A. Monschke, Elliott Ridgway, Ahmad Rushdi,
  Daniel Seidl, John Stephens, Laura~Painton Swiler, Anh Tran, and Justin
  Winokur.
\newblock Dakota, a multilevel parallel object-oriented framework for design
  optimization, parameter estimation, uncertainty quantification, and
  sensitivity analysis: Version 6.16 user's manual.

\bibitem{finkel2004convergence}
D~Finkel and Carl~Tim Kelley.
\newblock Convergence analysis of the direct algorithm.
\newblock Technical report, North Carolina State University. Center for
  Research in Scientific Computation, 2004.

\bibitem{zou2022neuraluq}
Zongren Zou, Xuhui Meng, Apostolos~F Psaros, and George~Em Karniadakis.
\newblock Neuraluq: A comprehensive library for uncertainty quantification in
  neural differential equations and operators.
\newblock {\em arXiv preprint arXiv:2208.11866}, 2022.

\bibitem{KOPRIVA1996244}
David~A. Kopriva and John~H. Kolias.
\newblock A conservative staggered-grid chebyshev multidomain method for
  compressible flows.
\newblock {\em Journal of Computational Physics}, 125(1):244--261, 1996.

\bibitem{PEYVAN2021110261}
Ahmad Peyvan, Jonathan Komperda, Dongru Li, Zia Ghiasi, and Farzad Mashayek.
\newblock Flux reconstruction using jacobi correction functions in
  discontinuous spectral element method.
\newblock {\em Journal of Computational Physics}, 435:110261, 2021.

\bibitem{kunz2003aerodynamics}
Peter~Josef Kunz.
\newblock {\em Aerodynamics and design for ultra-low Reynolds number flight}.
\newblock Stanford University, 2003.

\bibitem{2020SciPy-NMeth}
Pauli Virtanen, Ralf Gommers, Travis~E. Oliphant, Matt Haberland, Tyler Reddy,
  David Cournapeau, Evgeni Burovski, Pearu Peterson, Warren Weckesser, Jonathan
  Bright, St{\'e}fan~J. {van der Walt}, Matthew Brett, Joshua Wilson, K.~Jarrod
  Millman, Nikolay Mayorov, Andrew R.~J. Nelson, Eric Jones, Robert Kern, Eric
  Larson, C~J Carey, {\.I}lhan Polat, Yu~Feng, Eric~W. Moore, Jake
  {VanderPlas}, Denis Laxalde, Josef Perktold, Robert Cimrman, Ian Henriksen,
  E.~A. Quintero, Charles~R. Harris, Anne~M. Archibald, Ant{\^o}nio~H. Ribeiro,
  Fabian Pedregosa, Paul {van Mulbregt}, and {SciPy 1.0 Contributors}.
\newblock {{SciPy} 1.0: Fundamental Algorithms for Scientific Computing in
  Python}.
\newblock {\em Nature Methods}, 17:261--272, 2020.

\end{thebibliography}
